\documentclass[twocolumn,twocolappendix]{aastex631}

\usepackage{amsmath, amssymb, bm, booktabs, mathtools}

\begin{document}
\defcitealias{2025ApJ...985..265K}{Time Machine Design}

\title{Searches for Prompt Low-Frequency Radio Counterparts to Gravitational Wave Event S250206dm with the OVRO-LWA Time Machine}

\newcommand{\affilone}{Cahill Center for Astronomy and Astrophysics, California Institute of Technology, Pasadena, CA 91125, USA}
\newcommand{\affiltwo}{Owens Valley Radio Observatory, California Institute of Technology, Big Pine, CA 93513, USA}
\newcommand{\affilthree}{Department of Physics and Astronomy, Rice University, Houston, TX 77005, USA}
\newcommand{\affilfour}{Jet Propulsion Laboratory, California Institute of Technology, Pasadena, CA 91011, USA}
\newcommand{\affilfive}{School of Earth and Space Exploration, Arizona State University, Tempe, AZ 85287, USA}
\newcommand{\affilsix}{Center for Solar-Terrestrial Research, New Jersey Institute of Technology, Newark, NJ 07102, USA}
\newcommand{\affilseven}{George Mason University, Fairfax, VA 22030, USA}
\newcommand{\affileight}{University of New Mexico, Albuquerque, NM 87131, USA}
\newcommand{\affilnine}{Real-Time Radio Systems Ltd, Bournemouth, Dorset, BH6 3LU, UK}
\newcommand{\affilten}{Indian Institute of Technology Kanpur, Kanpur, Uttar Pradesh 208016, India}
\newcommand{\affilelev}{Rice Space Institute, Rice University, Houston, TX 77005, USA}

\newcommand{\affilassign}[1]{%
    \ifnum#1=1 \affiliation{\affilone}\fi
    \ifnum#1=2 \affiliation{\affiltwo}\fi
    \ifnum#1=3 \affiliation{\affilthree}\fi
    \ifnum#1=4 \affiliation{\affilfour}\fi
    \ifnum#1=5 \affiliation{\affilfive}\fi
    \ifnum#1=6 \affiliation{\affilsix}\fi
    \ifnum#1=7 \affiliation{\affilseven}\fi
    \ifnum#1=8 \affiliation{\affileight}\fi
    \ifnum#1=9 \affiliation{\affilnine}\fi
    \ifnum#1=10 \affiliation{\affilten}\fi
    \ifnum#1=11 \affiliation{\affilelev}\fi
}

\author[0000-0003-1226-118X]{Nikita Kosogorov}
\email{nakosogorov@gmail.com} 
\email{nkosogor@caltech.edu} 
\affilassign{1}
\affilassign{2}

\author[0000-0002-7083-4049]{Gregg Hallinan}
\affilassign{1}
\affilassign{2}

\author[0000-0002-4119-9963]{Casey Law}
\affilassign{1}
\affilassign{2}

\author{Jack Hickish}
\affilassign{9}

\author[0000-0003-1407-0141]{Jayce Dowell}
\affilassign{8}

\author{Kunal P. Mooley}
\affilassign{10}
\affilassign{1}

\author{Marin M. Anderson}
\affilassign{2}
\affilassign{4}

\author{Judd D. Bowman}
\affilassign{5}

\author{Ruby Byrne}
\affilassign{1}
\affilassign{2}

\author{Morgan Catha}
\affilassign{2}

\author{Bin Chen}
\affilassign{6}

\author{Xingyao Chen}
\affilassign{6}

\author{Sherry Chhabra}
\affilassign{6}
\affilassign{7}

\author{Larry D’Addario}
\affilassign{1}
\affilassign{2}

\author{Ivey Davis}
\affilassign{1}
\affilassign{2}

\author{Katherine Elder}
\affilassign{5}

\author{Dale Gary}
\affilassign{6}

\author{Charlie Harnach}
\affilassign{2}

\author{Greg Hellbourg}
\affilassign{1}
\affilassign{2}

\author{Rick Hobbs}
\affilassign{2}

\author{David Hodge}
\affilassign{1}

\author{Mark Hodges}
\affilassign{2}

\author{Yuping Huang}
\affilassign{1}
\affilassign{2}

\author{Andrea Isella}
\affilassign{3}
\affilassign{11}

\author{Daniel C. Jacobs}
\affilassign{5}

\author{Ghislain Kemby}
\affilassign{2}

\author{John T. Klinefelter}
\affilassign{2}

\author{Matthew Kolopanis}
\affilassign{5}

\author{James Lamb}
\affilassign{2}

\author{Nivedita Mahesh}
\affilassign{1}
\affilassign{2}

\author[0000-0002-2325-5298]{Surajit Mondal}
\affilassign{6}

\author{Brian O’Donnell}
\affilassign{6}

\author{Kathryn Plant}
\affilassign{2}
\affilassign{4}

\author{Corey Posner}
\affilassign{2}

\author{Travis Powell}
\affilassign{2}

\author{Vinand Prayag}
\affilassign{2}

\author{Andres Rizo}
\affilassign{2}

\author{Andrew Romero-Wolf}
\affilassign{4}

\author{Jun Shi}
\affilassign{1}

\author{Greg Taylor}
\affilassign{8}

\author{Jordan Trim}
\affilassign{2}

\author{Mike Virgin}
\affilassign{2}

\author[0000-0002-6611-2668]{Akshatha Vydula}
\affilassign{5}

\author{Sandy Weinreb}
\affilassign{1}

\author{Scott White}
\affilassign{2}

\author{David Woody}
\affilassign{2}

\author[0000-0003-2872-2614]{Sijie Yu}
\affilassign{6}

\author{Thomas Zentmeyer}
\affilassign{2}

\author[0000-0001-6855-5799]{Peijin Zhang}
\affilassign{6}



\begin{abstract}

We report on a search for prompt, low-frequency radio emission from the gravitational-wave (GW) merger \texttt{S250206dm} using the Owens Valley Radio Observatory Long Wavelength Array (OVRO-LWA). Early alerts favored a neutron-star–containing merger, making this a compelling target. Motivated by theoretical predictions of coherent radio bursts from mergers involving a neutron star, we utilized the OVRO-LWA \textit{Time Machine} system to analyze voltage data recorded around the time of the event. The \textit{Time Machine} is a two-stage voltage buffer and processing pipeline that continuously buffers raw data from all antennas across the array’s nearly full-hemisphere instantaneous field of view, enabling retrospective beamforming, dedispersion, and fast-transient candidate identification. For this event, we analyzed a 30-minute interval beginning 3.5 minutes after the merger, which included two minutes of pre-alert data recovered by the ring buffer. We searched the 50\% localization probability region with millisecond time resolution in the 69--86\,MHz frequency band. No radio counterpart was detected above a $7\sigma$ fluence detection threshold of $\sim$150 Jy\,ms. Using Bayesian analysis, we place a 95\% confidence upper limit on the source luminosity of $L_{95} = 4 \times 10^{41}\,\mathrm{erg\,s^{-1}}$. These constraints start to probe the bright end of the coherent-emission parameter space predicted by jet--ISM shock processes, magnetar and blitzar-like mechanisms, and recent simulation-based scenarios for neutron-star--containing mergers.
This study presents the first sensitive, large-area, millisecond-timescale search for prompt low-frequency radio emission from a GW merger with the OVRO-LWA, establishing a framework in which about ten additional events will yield stringent population-level constraints.

\end{abstract}


\section{Introduction}
\label{sec:intro}
Gravitational wave (GW) astronomy has opened new avenues for detecting compact binary mergers, including binary neutron star (BNS), neutron star–black hole (NSBH) and binary black hole (BBH) systems, enabling a multimessenger approach to studying these events.

The first observed BNS merger, GW170817 \citep{2017PhRvL.119p1101A,2017ApJ...848L..12A}, revealed a wealth of phenomena: it provided direct evidence for rapid neutron-capture (r-process) nucleosynthesis and heavy-element production via the optical/infrared counterpart known as a kilonova \citep[e.g.,][]{2017Sci...358.1556C,2017ApJ...848L..17C,2017Natur.551...67P,2017Natur.551...80K,2017Sci...358.1559K,2017Natur.551...75S}, confirmed the formation of relativistic jets in such mergers, thereby verifying a causal connection between BNS mergers and short gamma-ray bursts (GRBs) \citep[e.g.,][]{2017ApJ...848L..14G,2017ApJ...848L..15S,2017Sci...358.1565E,2017Sci...358.1579H,2018ApJ...856L..18M,2018ApJ...863L..18A,2018Natur.561..355M}, and constrained the viewing geometry of the system, which in turn improved the accuracy of the standard-siren measurement of the Hubble constant \citep[e.g.,][]{2017Natur.551...85A,2018Natur.561..355M,2018ApJ...863L..18A,2019NatAs...3..940H,2024PhRvD.109f3508P}. These discoveries were made possible through a broad range of electromagnetic observations, spanning the entire spectrum, including late-time radio detections days after the merger.

Despite these breakthroughs, one predicted counterpart remains elusive: prompt radio emission, occurring within milliseconds to minutes of the merger of BNS or NSBH systems \citep[e.g.,][]{2016MNRAS.459..121C,2019MNRAS.489.3316R,2025PhRvD.111h3023C}. Such emission could arise from several physical processes operating in the immediate pre- and post-merger phases, including but not limited to magnetic reconnection in the strongly magnetized pre-merger magnetospheres \citep[e.g.,][]{1996A&A...312..937L,2001MNRAS.322..695H,2012ApJ...755...80P,2016ApJ...822L...7W,2016MNRAS.461.4435M,2020ApJ...893L...6M,2023PhRvL.130x5201M,2023ApJ...956L..33M,2023MNRAS.519.3923C}, the activity of a newly formed hypermassive or supramassive neutron star remnant \citep[e.g.,][]{2011MNRAS.413.2031M,2013PASJ...65L..12T,2014A&A...562A.137F,2014MNRAS.441.2433R,2014ApJ...780L..21Z,2017ApJ...841...14M} and shocks when relativistic outflows interact with their environment. In the latter category, a strongly magnetized jet–ISM interaction can generate low-frequency radio pulses \citep{2000A&A...364..655U}, while accelerating binary-induced winds can self-shock and produce coherent radio precursors \citep{2021MNRAS.501.3184S}. Both analytic models and numerical simulations support the plausibility of these mechanisms \citep[e.g.,][]{2020ApJ...890L..24Z,2020ApJ...893L...6M}. Simulations in particular have shown that even moderately strong magnetic fields can generate fast radio burst (FRB)–like coherent transients near coalescence, with global magnetospheric models predicting emission via reconnection or shock-driven processes \citep[e.g.,][]{2020ApJ...893L...6M,2021MNRAS.501.3184S}. In addition to providing strong theoretical motivation, these studies suggest that a detection would offer a direct probe of the merger dynamics and local environment.

We note that although NSBH mergers are generally less likely to produce electromagnetic counterparts—because the neutron star often undergoes a nondisruptive plunge inside the black hole’s innermost stable circular orbit (ISCO), leaving negligible bound/ejected mass—certain configurations, particularly high black-hole spin and low mass ratios, can tidally disrupt the star outside the ISCO and leave disk/ejecta that power detectable emission, justifying continued observational efforts \citep[e.g.,][]{2015PhRvD..92d4028K,2021ApJ...923L...2F,2025PhRvD.111h3023C}. Even without disruption, General-Relativistic Magnetohydrodynamics (GRMHD) simulations indicate magnetospheric Poynting-flux outflows and a rapid post-merger “balding/monster-shock” phase; while the luminosity likely peaks at high energies, a coherent radio flash remains plausible \citep{East2021_BHNS_EM_noTidal,Kim2025_BH_pulsars_monster_shocks,2023ApJ...956L..33M}.

The ongoing O4 observing run of the LIGO (U.S.-based Laser Interferometer Gravitational-Wave Observatory), Virgo (European Gravitational Observatory in Italy), and KAGRA (Japan’s underground gravitational wave detector) network \citep[e.g.,][]{2004NIMPA.517..154A,2012JInst...7.3012A,2015CQGra..32b4001A,2016PhRvD..93k2004M,2019NatAs...3...35K,2021PTEP.2021eA101A,2025PhRvD.111f2002C}, which began in 2023, has already yielded numerous GW detections. In this study, we follow up on one such gravitational-wave event, \texttt{S250206dm}\footnote{\url{https://gracedb.ligo.org/superevents/S250206dm/view/}}, detected in February 2025 \citep[][]{2025GCN.39175....1L}. Based on its classification and inferred component masses, the event likely involved at least one neutron star.

The Owens Valley Radio Observatory Long Wavelength Array (OVRO-LWA) is a low-frequency radio interferometer situated at the Owens Valley Radio Observatory in California, at a geographic location of approximately $37.24^\circ$ N and $-118.28^\circ$ W. Although it utilizes the same basic dipole antenna design as the Long Wavelength Array stations in New Mexico \citep[e.g.,][]{2012JAI.....150004T}, OVRO-LWA was independently engineered to achieve all-sky imaging with arcminute-scale resolution, necessitating substantial custom hardware development. A recent expansion supported by the National Science Foundation’s Major Research Instrumentation (MRI) program (NSF AST-1828784) brought the total to 352 dual-polarization antennas. These include a dense central cluster of 241 elements within a $\sim$200 m footprint, complemented by 111 additional antennas deployed in an extended configuration reaching baselines of up to 2.4 km. The upgraded system offers continuous spectral coverage from 13 MHz to 86.5 MHz, divided into 24 kHz channels. Signals from all antennas are fully cross-correlated, enabling rapid, wide-field imaging across a $\sim$20,000 deg$^2$ field of view, with zenith resolution spanning from $\sim$5 to 30 arcminutes,corresponding to the highest and lowest frequencies of the band, respectively. Scientific applications of the array include exoplanet magnetospheres searches and stellar space weather monitoring \citep[e.g.,][]{2025ApJ...993...82D}, cosmological studies of the 21\,cm line from the Cosmic Dawn epoch \citep[e.g.,][]{2018AJ....156...32E,2019AJ....158...84E,2018MNRAS.478.4193P,2021MNRAS.506.5802G}, dynamic imaging spectroscopy of solar activity \citep[e.g.,][]{2021ApJ...906..132C,2025ApJ...992..143M,2025ApJ...992..128Z,2025ApJ...990L..50C,2025arXiv251012658M}, detection of extensive air showers from cosmic rays \citep[e.g.,][]{2020NIMPA.95363086M,2022icrc.confE.204P}, observations of meteor trail afterglows \citep[e.g.,][]{2024JGRA..12932272V}, investigations of galaxy clusters \citep[e.g.,][]{2023MNRAS.521.5786H}, low-frequency transient phenomena \citep[e.g.,][]{2018ApJ...864...22A,2019ApJ...886..123A}.

Prompt coherent radio searches encompass historical all-sky efforts, published rapid-response results, and forward-looking GW strategies. Historically, wide-FOV instruments at low frequencies below 1 GHz searched for prompt GRB emission and reported serendipitous constraints \citep[e.g.,][]{1975ApJ...196L..11B,1978Ap&SS..56..239I,1977Natur.267..815M,1981Ap&SS..75..153C,1982AIPC...77...79I,1998A&A...329...61B}, with high-frequency serendipity from COBE \citep{1997ApJ...487..114A}, and modern all-sky constraints from CHIME \citep{2023ApJ...954..154C,2024ApJ...972..125C}.  Published rapid-response radio results include low-frequency MWA follow-up 
of short/long GRBs \citep{2015ApJ...814L..25K,2021PASA...38...26A,2022PASA...39....3T,2022MNRAS.514.2756T,2025ApJ...982...32X}, 
LOFAR trigger campaigns on short/long GRBs 
\citep{2019MNRAS.490.3483R,2021MNRAS.506.5268R,2023MNRAS.526..106H,2024MNRAS.534.2592R,2025MNRAS.544...53H}, 
LWA1 foundational low-frequency prompt searches 
\citep{2014ApJ...785...27O,2012JAI.....150004T}, 
and ASKAP GHz-frequency searches for FRB-like bursts from Fermi GRBs 
\citep{2020MNRAS.497..125B}, 
together with historical rapid-trigger experiments (CLFST at 151\,MHz; 
FLIRT at 74\,MHz) \citep{1995Ap&SS.231..281G,1996MNRAS.281..977D,1998AIPC..428..585B} 
and complementary higher-frequency campaigns 
\citep{2012ApJ...757...38B,2014ApJ...790...63P}. In parallel, several works develop GW rapid-response strategies—for MWA \citep{2016PASA...33...50K,2019MNRAS.489L..75J,2023PASA...40...50T}, LOFAR \citep{2022MNRAS.509.5018G,2023MNRAS.523.4748G}, and ASKAP \citep{2020PASA...37...51W}. \citet{2020MNRAS.494.5110B} discuss late-time LOFAR follow-up of GW170817, including recommendations relevant for future rapid-response efforts. Within this context, OVRO–LWA has conducted early GRB/GW prompt searches \citep{2018ApJ...864...22A,2019ApJ...877L..39C}.

In this paper, we employ the \textit{Time Machine} system, developed for the OVRO-LWA and specifically designed for FRB-like searches of prompt radio emission associated with GW events. The complete design and validation of the system are detailed in \citet{2025ApJ...985..265K} (hereafter \citetalias{2025ApJ...985..265K}). Notably, the system uses a ring buffer that continuously stores digitized data, allowing capture beginning several minutes before the GW trigger and continuing for up to 30 minutes in total. In addition, the system supports offline beamforming—the process of digitally combining signals from antennas to synthesize narrow, steerable beams on the sky—as well as dedispersed transient searches.
Localization to sub-arcminute accuracy is then achieved through offline cross-correlation and imaging.

We apply the OVRO-LWA \textit{Time Machine} to the GW event \texttt{S250206dm}. In \autoref{sec:overview}, we describe the event properties and current multi-messenger constraints. \autoref{sec:obs} details the associated OVRO-LWA observations, data reduction and processing, and the optimization steps employed to search for candidate radio transients. In \autoref{sec:radio}, we develop the Bayesian framework used to constrain the properties of radio emission, and in \autoref{sec:discussion}, we discuss the broader implications of our results. Finally, \autoref{sec:sum} summarizes our findings.

\section{S250206dm overview}
\label{sec:overview}

\begin{figure*}[htbp]
  \centering
  \includegraphics[width=\textwidth]{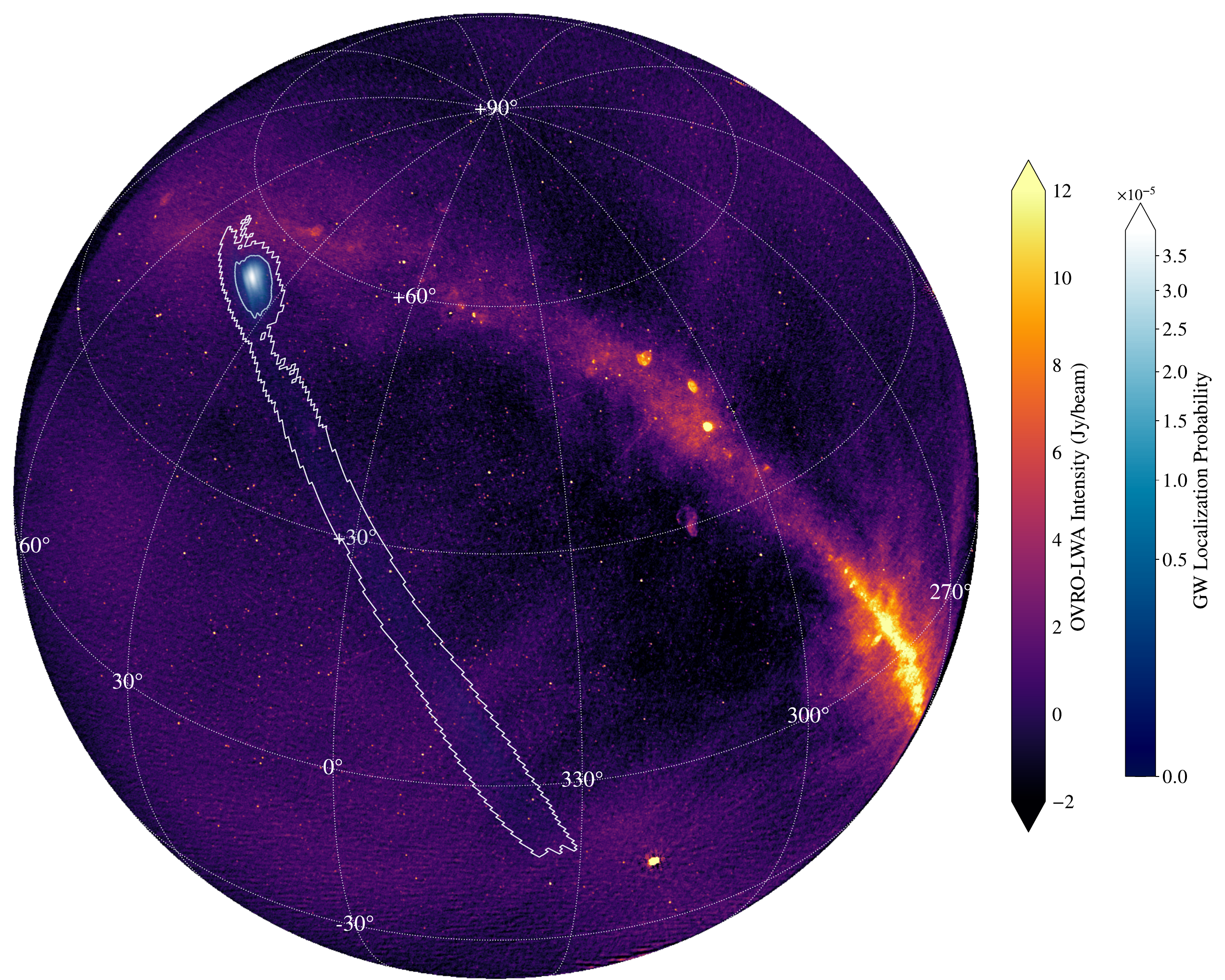}
  \caption{All-sky OVRO-LWA map generated from online cross-correlation data for a single 10\,s integration following the merger, averaged over the 55--86\,MHz band. The GW localization probability is overlaid, with the white contours marking the 50\% and 90\% credible regions. The Sun is visible near $\mathrm{RA}\approx321^{\circ},\ \mathrm{Dec}\approx-15^{\circ}$.}

  \label{fig:allsky_localization}
\end{figure*}

\subsection{S250206dm Properties}
\label{sec:ligo_property}

Candidate merger \texttt{S250206dm} occurred at 2025-02-06 21{:}25{:}30~UTC, as reported by the LIGO–Virgo–KAGRA Collaboration. The false-alarm rate (FAR) for this event is approximately one per 25~years. The first alert was issued at 21\text{:}26\text{:}02~UTC by the MBTA pipeline \citep[e.g.,][]{2016CQGra..33q5012A} as a preliminary notice, reporting a FAR of approximately one per 1.2~years. Approximately five minutes later, at 21\text{:}30\text{:}51~UTC, the PyCBC pipeline \citep[e.g.,][]{2018PhRvD..98b4050N} issued a second preliminary alert with a significantly lower FAR of one per 25~years, which triggered our buffer.

Subsequent analyses strongly suggest a NSBH or BNS merger, with classification probabilities of NSBH~55\% and BNS~37\%, and a high probability of at least one neutron star being present (HasNS~$=$100\%). The probability of matter remaining outside the final compact object (\texttt{HasRemnant}) is estimated at approximately $30\%$. The 90\% credible region for the sky localization initially spanned $2139~\mathrm{deg}^2$ but was later refined to $547~\mathrm{deg}^2$ in the final sky map. The luminosity distance was similarly improved from a broader initial estimate to a final value of $373 \pm 104$~Mpc, reported approximately 1.5~days after the first alert \citep[e.g.,][]{2016PhRvD..93b4013S,2015PhRvD..91d2003V}.

We note that a majority of the probability density lay above the horizon for OVRO-LWA at the time of the merger. The probability distribution over the visible sky at that moment is shown in \autoref{fig:allsky_localization}.

\subsection{Current Multi-messenger Constraints on S250206dm}

Despite an extensive multi-wavelength follow-up campaign by the astronomical community, no electromagnetic or neutrino counterpart to \texttt{S250206dm} has been confirmed.

High-energy searches by gamma-ray and X-ray observatories yielded no prompt emission or afterglow, placing stringent upper limits on any associated short gamma-ray burst. A massive optical and near-infrared campaign was undertaken by numerous facilities. This effort identified several transient candidates, though subsequent photometric and spectroscopic follow-up revealed them to be unrelated phenomena or artifacts \citep{2025arXiv250700357A, 2025PASP..137g4203F,2025ApJ...990L..46H, 2025GCN.39210....1Y}.

Two other potential associations drew initial interest: a pair of track-like neutrino events reported by IceCube \citep{2025GCN.39176....1I} and a fast radio burst (FRB~20250206A) detected by CHIME \citep{2025GCN.39216....1C}. In both cases, the refined GW sky localization showed a significant spatial disagreement with these events, leading to the conclusion that they were likely chance coincidences.

\section{OVRO-LWA Observations and Data Processing} \label{sec:obs}

We refer the reader to the \citetalias{2025ApJ...985..265K} paper for a detailed description of the OVRO-LWA GW follow-up system. Below, we summarize the key components, highlight the modifications specific to the search for \texttt{S250206dm}, and provide a comprehensive description of the analysis as applied to this event.

\subsection{Ring Buffer Triggering} 
\label{sec:trigger}

We note that the first preliminary alert for \texttt{S250206dm} did not meet our selection criterion of a FAR lower than one per 10 years (see \autoref{sec:ligo_property}). However, a subsequent notice with a significantly improved FAR successfully satisfied our triggering conditions; this PyCBC alert was issued at 21\text{:}30\text{:}51~UTC (5~min~21~s post-merger). This prompted the system to save data from its ring buffer, with the recorded observation commencing at 21\text{:}28\text{:}59~UTC (3~min~29~s post-merger), demonstrating that the buffer recovered 1~min~52~s of pre-alert data relative to the triggering notice. Accordingly, the observational latency is largely set by the LIGO--Virgo--KAGRA (LVK) alert chain rather than the OVRO-LWA on-site response.

The saved voltage buffer spans 30\,min and covers 55--86\,MHz. For the prompt-search analysis, we process the 69--86\,MHz sub-band and perform offline beamforming to form tied-array beams with 1.3\,ms temporal and 0.7\,kHz spectral resolution (see \autoref{sec:beamform}), followed by an incoherent dedispersion search (\autoref{sec:dedispersion}). Simultaneous cross-correlation visibilities were also recorded for calibration (\autoref{sec:cross-corr}).

Raw data were transferred to the 5-PB storage system of the OVRO-LWA within approximately three hours, after which the \textit{Time Machine} real-time monitoring operations resumed. We note that the system was operational for most of the O4 observing run to date, with rare interruptions for maintenance and occasional tests using BBH events.

Finally, the 3~min~29~s gap between the merger and the recording start can be limiting at the low end of the dispersion measure (DM) range but is acceptable for moderate–to–high DMs within our search. Specifically, the cold-plasma dispersion delay to our band is \(\sim0.9\)–\(3.7\) min at 86\,MHz and \(\sim1.5\)–\(5.8\) min at 69\,MHz for \(\mathrm{DM}=100\)–\(400~\mathrm{pc\,cm^{-3}}\) (see \autoref{sec:dedispersion}). Thus, for DMs toward the upper half of this range, merger-coincident emission would naturally arrive minutes after the GW trigger. Such low-frequency dispersion delays have been discussed in earlier work (e.g., \citealt{1997NewA....2..555L}), sometimes referred to as a “radio time-machine’’ effect. Buffering further increases capture probability. Scattering can also broaden and slightly delay the signal, but it is expected to be much smaller than the dispersion delay and does not change this conclusion; if present, it would only shift additional power to later times.

\subsection{Simultaneously Collected Cross-correlation Data} \label{sec:cross-corr}

\begin{figure*}[t]
    \centering
    \includegraphics[width=\textwidth]{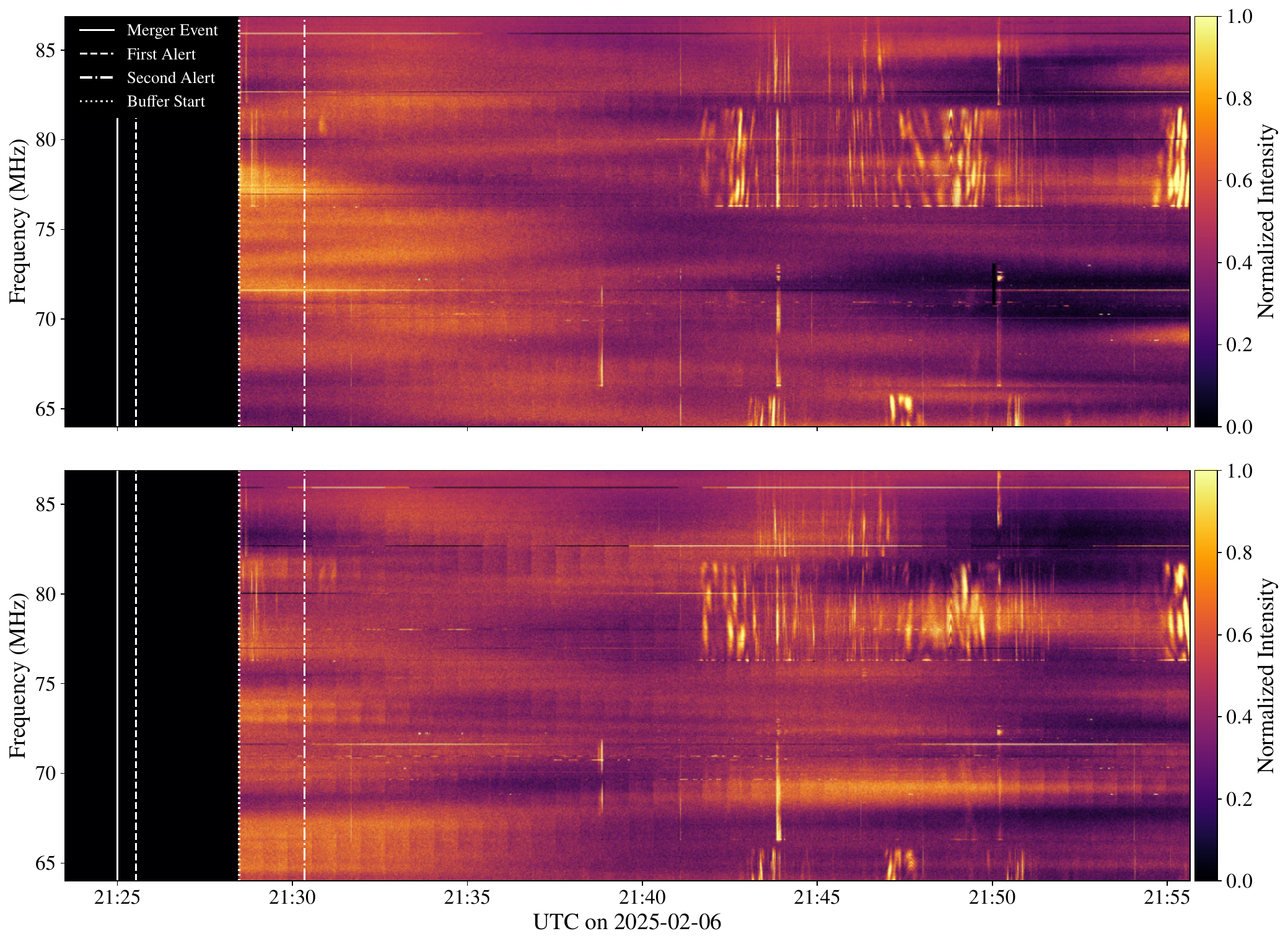}
    \caption{Normalized dynamic spectra (waterfall plots) from two offline-formed OVRO-LWA beams, centered at J2000 $\mathrm{RA}=30.87^{\circ},\ \mathrm{Dec}=+51.98^{\circ}$ (top) and $\mathrm{RA}=338.70^{\circ},\ \mathrm{Dec}=+12.17^{\circ}$ (bottom). For visualization, the raw data ($>2\times10^{4}$ channels and $>10^{6}$ time samples) were averaged in time and frequency. Vertical guide lines mark key times: the merger epoch (solid), the first and second preliminary GW alerts (dashed, dash--dotted), and the start of the buffer (dotted). To enhance faint, time-variable structure, a time-median bandpass was subtracted from each integration; the result was then min--max normalized using the 0.5th and 99.5th percentiles and clipped to $[0,1]$. Visible features include: thin, persistent horizontal lines at fixed frequencies that are consistent with narrowband terrestrial carriers such as legacy VHF-low TV channels; brief band-wide brightenings that appear as near-vertical stripes in both beams, likely due to broadband impulsive RFI or occasional instrumental gain changes; and brief, few-channel brightenings consistent with meteor or aircraft scatter of narrowband signals. Slow background variations differ between beams, likely reflecting sky brightness, sidelobe pickup from bright sources, and ionospheric effects.
    }
    \label{fig:waterfall}
\end{figure*}

Before proceeding with the offline beamforming and transient searches, we derived complex, frequency-dependent gain solutions from a separate, continuously recorded online cross-correlation dataset. This system runs in parallel with the triggered voltage buffer and continuously saves 10\,s integrated visibility data, which are buffered for several days. This allowed us to retain the 10\,s-cadence data from the entire period, including both pre-merger time and the later calibrator transit. This dataset is ideal for calibration and all-sky imaging at 10\,s time resolution but is not used for the millisecond-timescale transient search itself. We first flagged antennas showing excessive temporal variability, abnormally low power maxima, or persistently high power minima and anomalous spectral shapes; in total, fewer than 3\% of antennas were removed. We additionally flagged radio frequency interference (RFI) using the \texttt{AOFlagger} software\footnote{\href{https://gitlab.com/aroffringa/aoflagger}{https://gitlab.com/aroffringa/aoflagger}} \citep{2010ascl.soft10017O}.

A simplified sky model—using flux densities and spectral indices for the brightest radio sources \citep[][]{1977A&A....61...99B,2017ApJS..230....7P}—was employed to set the absolute gains. Using this online cross-correlation dataset, we selected a 4-minute interval around the transit of the calibrator source Cassiopeia A (occurring at $\sim$22:07 UTC) and solved for the bandpass using the \texttt{bandpass} task within the Common Astronomy Software Applications (CASA)\footnote{\url{https://casa.nrao.edu/}} \citep{2022PASP..134k4501C}. The resulting per-antenna amplitude and phase solutions were smooth and well-behaved.

To confirm the quality of the solutions, imaging and deconvolution were performed with \texttt{WSClean}\footnote{\url{https://gitlab.com/aroffringa/wsclean}} \citep{2014MNRAS.444..606O} over the visible hemisphere on a $4096 \times 4096$ grid. We also peeled several bright sources using the \texttt{TTCal} software\footnote{\href{https://github.com/mweastwood/TTCal.jl}{https://github.com/mweastwood/TTCal.jl}}\citep{2016zndo...1049160E}. The Sun was left unpeeled in the illustrative all-sky image, as it lies well outside the region of interest. Crucially, this entire imaging and peeling process was only for visualization and does not affect the transient search, which operates directly on the separate, beamformed data.

In \autoref{fig:allsky_localization}, we present an image generated from the online cross-correlation data, using the procedure described above, to visually illustrate the full sky and the localization of the GW event. The image integrates data over a 10\,s interval starting from the merger time and spans the 55–86\,MHz band. We applied Briggs weighting \citep[][]{1995AAS...18711202B} with a robust parameter of 0.

\subsection{Offline Beamforming} 
\label{sec:beamform}

Following calibration, we transform the buffered voltages into a set of coherent, direction-dependent time series in order to search for millisecond-duration transients associated with the GW event. Specifically, we perform offline beamforming using the core-only OVRO-LWA antennas to optimize sensitivity while maintaining computational efficiency, applying the derived gains. The beamformed data have a time resolution of 1.3 ms. Example waterfall plots from two beams are shown in \autoref{fig:waterfall}.

A brief summary of the beamformed observation parameters is as follows: processed voltages are upchannelized to 0.7\,kHz to limit dispersion smearing for DM of $\lesssim 400~\mathrm{pc\,cm^{-3}}$, then multiplied by the corresponding gains derived from cross-correlation data and phased toward a grid of directions covering the accessible, high-probability region of the GW event localization. The offline beamforming pipeline was developed using the Bifrost framework \citep{2017JAI.....650007C}, incorporating newly developed Bifrost blocks\footnote{\url{https://github.com/realtimeradio/caltech-bifrost-dsp}}.

Due to computational constraints of the transient search (see \autoref{sec:dedispersion} for details), we restricted our analysis to the 50\% localization probability region of the GW sky map, ensuring full coverage of this area with 51 overlapping beams. Additionally, we formed a beam toward the sky position of the CHIME detected event FRB20250206A, which occurred within minutes of the merger and was considered a potential spatially or temporally coincident transient.

Finally, similar to the testing presented in \citetalias{2025ApJ...985..265K}, we verified the sensitivity and data quality of our beamformer using observations of the pulsar PSR~B1919+21, which was successfully detected.

\subsection{Prompt Radio Pulse Search}
\label{sec:dedispersion}

\subsubsection{Dedispersion Strategy}
\label{sec:dispersion_plan}

\begin{figure*}[t]
    \centering
    \includegraphics[width=\textwidth]{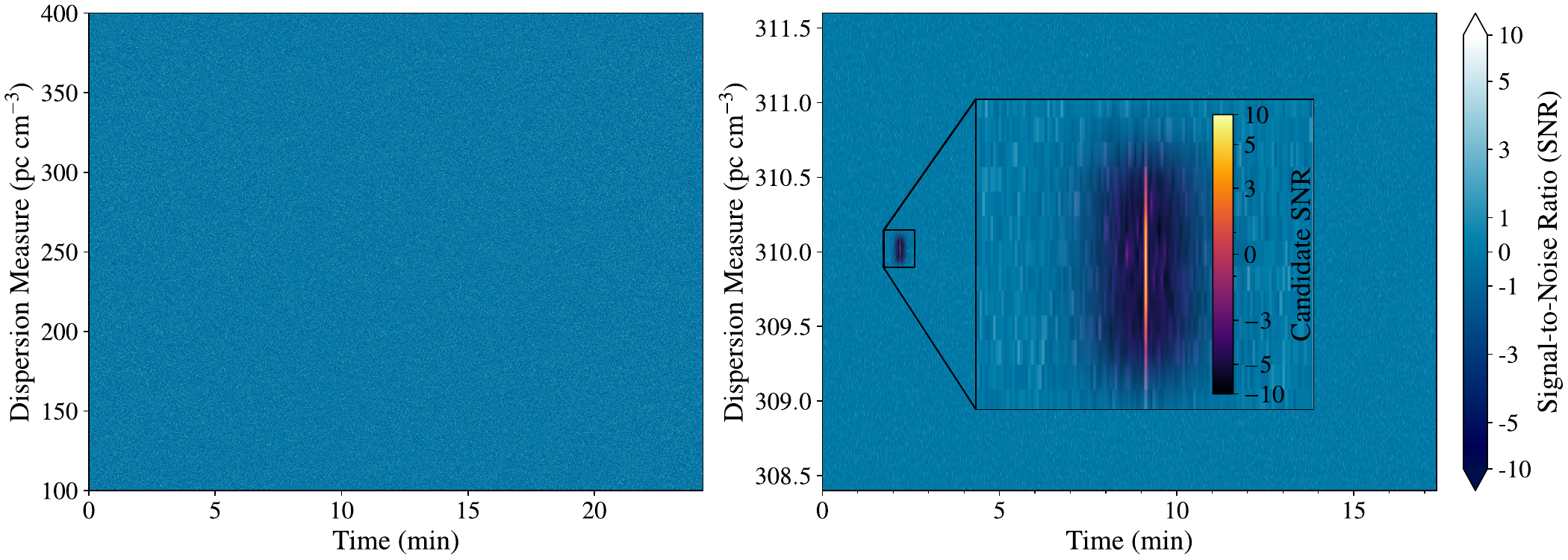}
    \caption{
        Left: Signal-to-noise ratio (SNR) plotted over dispersion measure DM and time since the start of the observation (in minutes), for the same beam shown in the top panel of \autoref{fig:waterfall}.  
        Right: Similar plot for the beam with an injected signal. Pixels with $\mathrm{SNR} > 7$ are overlaid using a semi-transparent colormap with slight dilation and blurring to enhance visual contrast. The shared color bar indicates SNR values for both panels, and the inset highlights the peak detection region. The slight negative bowl around the injected signal is purely from the visualization baseline subtraction and is not astrophysical. 
    }
    \label{fig:dedispersion}
\end{figure*}

A dispersion plan was developed to optimally sample the relevant DM range while minimizing computational demands. We note that the dedispersion search must be performed with sufficiently small DM steps to avoid smearing and missing any potential signal whose true DM lies between trial values. A brute-force incoherent dedispersion over the full 55–85\,MHz band at millisecond time resolution for DMs ranging from 0 to 400\,pc\,cm$^{-3}$ would require extremely fine DM steps, resulting in more than $10^5$ trials. This would impose prohibitive computational and data-storage demands.

The maximum acceptable dispersion smearing per trial is set by the condition
\[
\Delta t_{\rm smear} \propto \frac{\Delta\mathrm{DM} \times \mathrm{BW}}{f_{\rm ctr}^3} \lesssim \Delta t_{\rm samp},
\]
where $\mathrm{BW}$ is the bandwidth, $\Delta\mathrm{DM}$ is the step size in dispersion measure, $f_{\rm ctr}$ is the central frequency, and $\Delta t_{\rm samp}$ is the sampling time. This scaling relation implies that larger bandwidths and lower central frequencies require finer DM steps to avoid excessive smearing.

To mitigate this issue, we restricted the search band to 69--86\,MHz, thereby reducing the effective bandwidth and relaxing the DM step requirement. Additionally, we limited the DM search range to 100--400\,pc\,cm$^{-3}$. This range was chosen conservatively: the lower bound reflects the minimum Galactic contribution, while the upper limit of $400~\mathrm{pc\,cm^{-3}}$ (see \citetalias{2025ApJ...985..265K} for more discussion) bracketed the plausible additional DM from the IGM, host galaxy, and immediate merger environment, which are not tightly constrained by the LVK distance alone. These choices together reduced the total number of dedispersion trials by more than a factor of two.

The effective time resolution is set by adding in quadrature the contributions from dispersion smearing, residual DM-step error, and intra-channel filter response \citep[e.g.,][]{2003ApJ...596.1142C,2023RvMP...95c5005Z}. These three terms alone remain at only a few milliseconds, which already provides essentially full sensitivity for the fast-transient parameter space targeted by our search.

\subsubsection{Search Implementation}
\label{sec:pipeline_processing}

We use \texttt{PyTorchDedispersion}\footnote{\url{https://github.com/nkosogor/PyTorchDedispersion}}, built on the PyTorch library\footnote{\url{https://pytorch.org/}} \citep[][]{2019arXiv191201703P}, which enables Graphics Processing Unit (GPU)-accelerated searches for dispersed radio transients. The software performs incoherent dedispersion, applies boxcar filtering with a range of widths, removes slow trends in the data, and identifies candidate signals based on predefined Signal-to-noise ratio (SNR) thresholds. To mitigate the effects of radio frequency interference (RFI), we identify and exclude bad channels using Spectral Kurtosis (SK) and Savitzky--Golay filtering \citep[e.g.,][]{1964AnaCh..36.1627S,2010PASP..122..595N}.

In order to search for prompt radio pulses of varying temporal widths, we applied boxcar filters with widths from 1\,ms to 200\,ms (in 1\,ms increments) to the dedispersed time series, as this step is computationally inexpensive relative to the dedispersion itself. The maximum boxcar width is motivated by expected scattering timescales (see \citetalias{2025ApJ...985..265K}). All peaks exceeding an SNR threshold of 5.0 across any trial DM are recorded.

This full search procedure was applied independently to all 51 beams covering the 50\% localization region, and one beam toward FRB20250206. Example results for one dedispersed beam are shown in \autoref{fig:dedispersion}. We further validate the system using injection tests, following the methodology outlined in the design paper. The same figure also presents an example of an injected high SNR signal.

\subsection{Investigation of Cloud Computing for Pipeline Profiling and Optimization}
\label{sec:cloud}
We note that even with a reduced number of DM trials, dedispersion remains computationally intensive. This is primarily due to the volume and resolution of the data: each 30-minute observation contains over one million time steps and tens of thousands of frequency channels across the 69--86\,MHz band. For each beam, the dedispersion process must evaluate tens of thousands of trial DMs, effectively constructing a three-dimensional data cube.  
The resulting memory footprint exceeds the capacity of the GPUs available for offline processing, which in our case are 16 NVIDIA RTX A4000 graphics cards with 16 GB of memory each. To address this, we adopt a tiling strategy that partitions the data into smaller, memory-efficient chunks for sequential processing. Nonetheless, our current implementation still requires over a month to process all beams. To make the analysis computationally feasible, we therefore restricted it to the 50\% localization region of the GW sky map.

To improve performance, we explored cloud-based optimizations. Cloud instances offer scalable compute resources and, importantly, provide isolated environments that are not affected by resource contention or background activity on the shared OVRO-LWA servers. This isolation allows for more consistent and reliable GPU performance, which is important for our dedispersion pipeline testing. Specifically, we recreated our entire dedispersion software stack on an Amazon Web Services (AWS) spot instance\footnote{\url{https://aws.amazon.com/ec2/spot/}} (\texttt{g4dn.xlarge}, T4 GPU, CUDA~12.6\footnote{\url{https://developer.nvidia.com/cuda-toolkit}}). Test data for one beam were initially uploaded to Amazon S3\footnote{\url{https://aws.amazon.com/s3/}} and retrieved by the instance as needed.

Our first phase of optimization focused on improving hardware utilization. We re-architected the pipeline into a producer-consumer model, using a background thread to prefetch data chunks and two CUDA streams to overlap data transfer and computation. By switching to page-locked (pinned) host memory and reusing pre-allocated GPU buffers, we doubled the host-to-device transfer rate and boosted average GPU occupancy above 50\%. These implemented changes yielded a ~10–15\% increase in overall throughput.

Nonetheless, further profiling revealed additional fundamental bottlenecks in the existing implementation. Disk I/O remains CPU-bound, and the beam-data layout—though contiguous—is not optimized for our access patterns. In PyTorch, even with optimizations, memory-access patterns are abstracted, preventing fine control of shared and register memory. As a result, implementing custom CUDA C++ kernels with explicit shared-memory management could yield further gains. A detailed exploration of these solutions is beyond the scope of this paper and will be pursued in future work.

In summary, the offline pipeline is sufficiently developed to enable large-scale analyses, but fundamental computational bottlenecks still limit performance. Consequently, the restriction of the search band to 69--86\,MHz and the analysis to the 50\% localization region reflects computational necessity rather than astrophysical motivation. Continued optimization will be essential to reduce turnaround times and expand sky coverage in future searches.

\subsection{Candidate Identification} \label{sec:cand}
To distinguish genuine astrophysical signals from statistical fluctuations, we clustered surviving triggers using the DBSCAN algorithm \citep[][]{1996kddm.conf..226E}. This method identifies dense groups of detections in the time–DM-width parameter space—consistent with real dispersed pulses—while classifying isolated triggers as noise. Each resulting cluster was treated as a single candidate event and ranked by its peak SNR. We adopt a 7-sigma detection threshold, which corresponds to of order unity expected false alarms across our search space of over a million time steps, tens of thousands of DM trials, and up to a couple of hundred boxcar widths—i.e., over $10^{12}$ statistical trials. This threshold is derived assuming Gaussian noise and statistical independence between trials, where the tail probability above 7 sigma is $\sim 1.3 \times 10^{-12}$. In practice, the trials are correlated, reducing the effective number of independent trials. Thus, the 7-sigma threshold serves as a conservative choice, limiting false positives while preserving sensitivity to bright signals. This $7\sigma$ threshold corresponds to a fluence sensitivity 
of $\sim150$\,Jy\,ms. We report that no candidates above the threshold passed the selection criteria.

\begin{figure}[h!]
    \centering
    \includegraphics[width=0.478\textwidth]{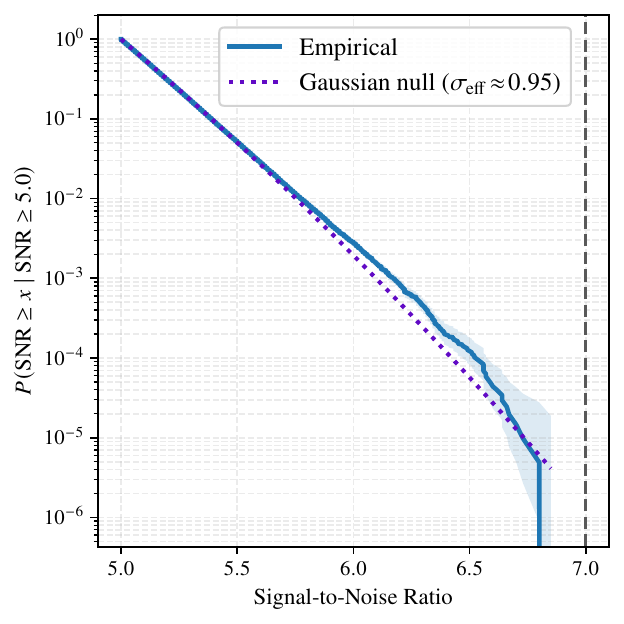}
    \caption{Complementary cumulative SNR distribution of surviving triggers with
    $\mathrm{SNR}\ge5$.  The blue curve is the empirical survival probability with
    a 95\% binomial band (shaded).  The dotted purple curve is a conditional Gaussian fit with effective width $\sigma_{\rm eff}\approx0.95$ that accounts for the SNR cut and mild correlations.  The vertical dashed line marks the adopted $7\sigma$ detection threshold; no triggers exceed this level.}
    \label{fig:snr_survive}
\end{figure}

To illustrate the statistical quality of the surviving triggers, \autoref{fig:snr_survive} shows the complementary cumulative SNR distribution (survival function) for events with $\mathrm{SNR}\ge5$, together with a conditional Gaussian fit.  The close agreement, with an effective width $\sigma_{\rm eff}\approx0.95$, confirms that residual noise behaves nearly Gaussian and that the absence of $7\sigma$ candidates is consistent with the expected false-alarm rate.

We further formed a beam toward the CHIME-reported FRB~20250206A position and searched with trial DMs bracketing the reported value ($\mathrm{DM}=207.117\pm0.003~\mathrm{pc\,cm^{-3}}$) and boxcar widths of 1–200~ms; no candidate above our $7\sigma$ threshold was found, consistent with the sub-ms characteristic scattering timescale ($\sim$0.05–0.10~ms at 600~MHz) \citep{2025GCN.39216....1C}.

\section{
Bayesian Inference Framework for S250206dm}
\label{sec:radio}

Given the absence of a detected prompt radio counterpart, we develop a Bayesian
inference framework to combine the OVRO--LWA radio observations with external
astrophysical information for the compact-object merger \texttt{S250206dm}. 
This approach quantifies how the GW-derived distance and sky localization, 
physically motivated models for the Milky Way, intergalactic media, and host DM 
contributions, and the radio non-detection together constrain the physical
properties of the merger environment and any potential prompt radio emission.
In this hierarchical model, the LVK three-dimensional posterior on sky position
and distance and the DM models enter as priors, while the OVRO--LWA dataset
provides the likelihood. In the case of a detection, the framework enables joint
inference on both astrophysical and propagation parameters, improving the
posteriors on quantities such as distance and DM components; in the case of a
non-detection, it yields upper limits on intrinsic quantities such as the radio
luminosity.

\begin{figure*}[t]
  \centering
  \includegraphics[width=\textwidth]{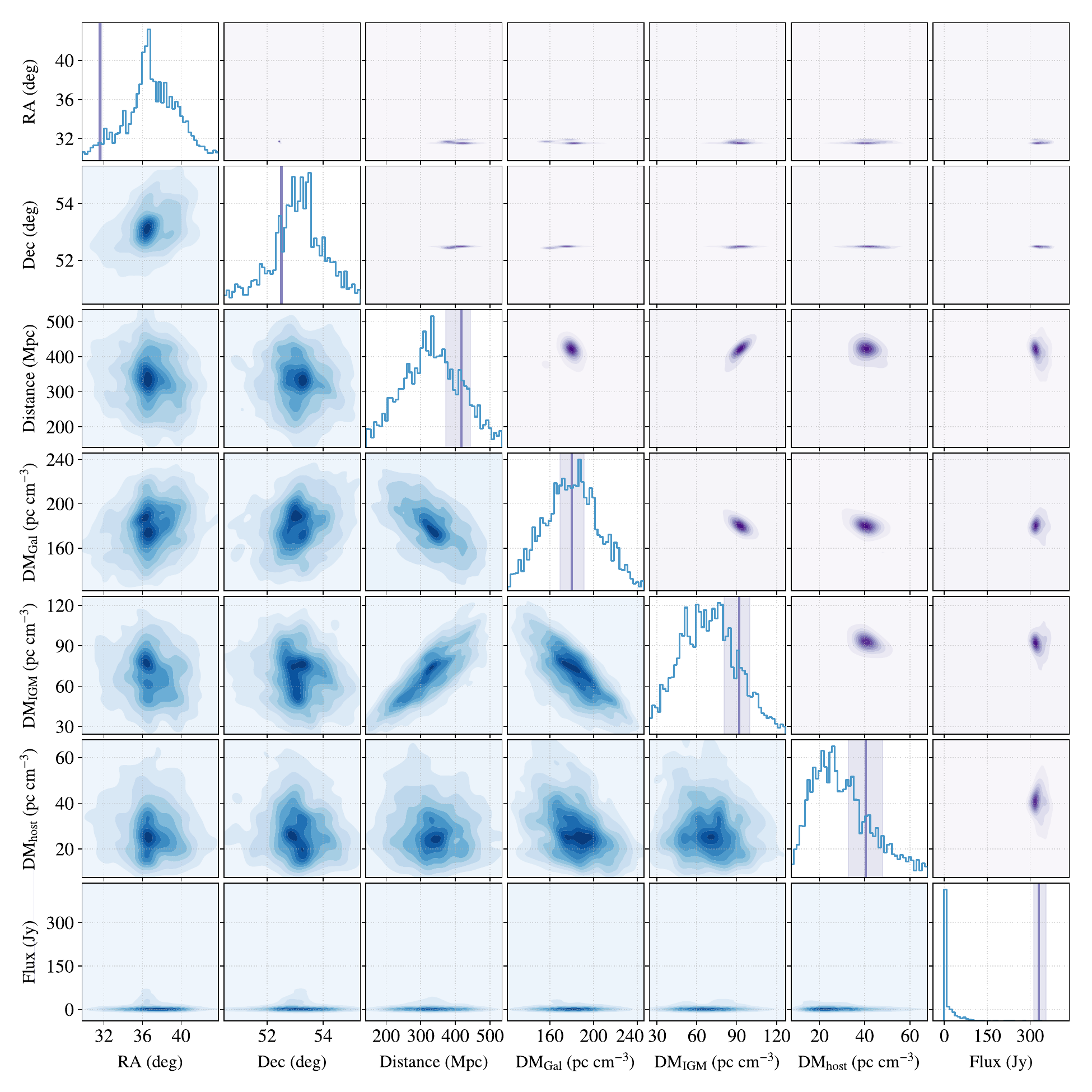}
  \caption{
    Comparison of the posterior distributions for an injected detection
    (upper triangle, purple color) versus a non-detection
    (lower triangle, blue color).  Diagonal panels show the marginal
    histograms of the non-detection posterior (blue lines) overlaid with
    the 16–84\% credible intervals and medians of the detection posterior
    (purple spans and lines).  Off-diagonal panels illustrate the joint
    posteriors via kernel density estimates for each case.  From left to
    right (and top to bottom), the parameters are:
    RA, Dec, Distance, $\mathrm{DM}_{\mathrm{Gal}}$,
    $\mathrm{DM}_{\mathrm{IGM}}$, $\mathrm{DM}_{\mathrm{host}}$ and Flux density.
  }
  \label{fig:detection_vs_nondetection}
\end{figure*}

\subsection{Bayesian Analysis Framework}
We aim to compute the posterior probability distribution of the model parameters
$\theta = \{i, D, \mathrm{DM}, t_0, F\}$ by applying Bayes' theorem,
\[
p(\theta \mid \mathcal{D}) = \frac{p(\mathcal{D} \mid \theta)\,p(\theta)}{p(\mathcal{D})},
\]
where $\mathcal{D}$ denotes the complete radio observation dataset for the candidate event (see \autoref{sec:forward}), $p(\mathcal{D} \mid \theta)$ is the likelihood of observing the data given the parameters, $p(\theta)$ is the prior encoding astrophysical knowledge, and $p(\mathcal{D})$ is the Bayesian evidence (a normalization constant in this context). This framework combines the data with prior expectations under a hierarchical model.

The parameter set $\theta$ includes the sky pixel $i$, distance to the source $D$, total dispersion measure $\mathrm{DM}$, arrival time $t_0$, and intrinsic flux density $F$. Here $i$ labels a HEALPix sky pixel on the LVK BAYESTAR skymap (i.e., a sky 
position $\boldsymbol{\Omega}$), and $D$ is the corresponding GW-inferred luminosity distance along that line of sight.  
While a more general model could also include the intrinsic pulse width $w$ and the scattering timescale $\tau_{\mathrm{sc}}$, these quantities mainly describe signal attenuation due to temporal scattering and the use of a specific boxcar filter width. Because our search pipeline tests a wide range of boxcar widths from 1 to 200\,ms (see \autoref{sec:dedispersion}), any detectable event is expected to be well matched to at least one trial width, minimizing attenuation. To simplify the model and avoid imposing poorly constrained priors on scattering, we therefore omit $w$ and $\tau_{\mathrm{sc}}$ from the Bayesian inference.

For each tied beam $b = 1, \dots, B$, we record a millisecond-resolution time series covering the full observation duration. After dedispersion we obtain a flux-density estimator $\widehat{F}_b$ in Jy. We collect the estimators from all $B$ beams into the data vector $\mathbf{d}_\mathrm{obs} = \bigl(\widehat{F}_1, \dots, \widehat{F}_B\bigr)^{\!\mathsf{T}}$.

\subsection{Forward Model and Priors}
\label{sec:forward}
We note that \( \mathcal{D} = \{\mathbf{d}_\mathrm{obs}, \mathrm{DM}_\mathrm{obs}, t_{0,\mathrm{obs}} \} \) represents the full radio observation dataset for the event, $\mathrm{DM}_\mathrm{obs}$ the best‐fit trial DM, and $t_{0,\mathrm{obs}}$ the best‐fit arrival time. The total likelihood naturally separates into three components:
\[
p(\mathcal D\mid\theta)
= p(\mathbf{d}_\mathrm{obs}\mid\theta) \times
  p(\mathrm{DM}_\mathrm{obs}\mid\theta) \times
  p(t_{0,\mathrm{obs}}\mid\theta).
\]

The first term, \( p(\mathbf{d}_\mathrm{obs} \mid \theta) \), represents the likelihood of the beamformed flux data given the source parameters. The second term, \(p(\mathrm{DM}_\mathrm{obs}\mid\theta)\ \), captures the consistency between the trial dispersion measure and the predicted value from the DM model. The third term, \( p(t_{0,\mathrm{obs}}\mid\theta) \), quantifies the alignment between the trial arrival time and the model-predicted arrival time. Both the DM and time likelihoods are modeled as Gaussians centered at the predicted values \( \mathrm{DM} \) and \( t_0 \), reflecting the assumption that for high-SNR events the trial parameters closely recover the true ones.

The forward model predicts the observed data vector $\mathbf{d}_\mathrm{obs}$ for a given set of source parameters. The model is expressed as:
\[
\mathbf{d}_\mathrm{obs} = F\,\mathbf{r}(\boldsymbol{\Omega}) + \boldsymbol{n},
\]
where $\mathbf{r}$ is the response vector accounting for the telescope's intrinsic sensitivity pattern, and $\boldsymbol{n}$ is a noise vector, which we model as a zero-mean multivariate Gaussian whose covariance matrix, $\boldsymbol{\Sigma}_{\!n}$, is measured from noisy beamformed data. We include this covariance matrix since overlapping beams are not fully independent.

Given this model, the likelihood of observing the data $\mathbf{d}_\mathrm{obs}$ is a multivariate Gaussian:
\[
    p(\mathbf{d}_\mathrm{obs}|\theta) = \mathcal{N}_{B}(\mathbf{d}_\mathrm{obs} \,;\, F\mathbf{r}, \boldsymbol{\Sigma}_{\!n}).
    \label{eq:fulllikelihood}
\]

The framework adopts physically motivated, hierarchical priors, as summarized in Table~\ref{tab:priors}. In particular, the prior on sky position $i$ and distance $D$ is derived from the BAYESTAR skymap \citep[][]{2016PhRvD..93b4013S}, which provides the joint distribution $p(i, D) = p(D|i)\,p(i)$. The functional form and parameterization of the corresponding probability density functions (PDFs) are described in \citet{2016ApJ...829L..15S, 2016ApJS..226...10S}.

\begin{table}[h!]
\centering
\caption{Summary of Priors for the Bayesian Analysis.}
\label{tab:priors}
\begin{tabular}{@{}ll@{}}
\toprule
\textbf{Parameter} & \textbf{Prior Distribution} \\
\midrule
Sky Pixel $i$ & From the GW event's skymap, $p(i)$ \\
Distance $D$ & Conditional, from the skymap, $p(D|i)$ \\
Arrival Time $t_0$ & Uniform over observation time $\mathcal{T}$ \\
Milky Way DM & $\mathcal{N}(\mu_{\mathrm{Gal},i},\, [0.2\,\mu_{\mathrm{Gal},i}]^2)$ (YMW16 model) \\
IGM DM & $p\!\big(\mathrm{DM}_{\rm IGM}\mid D\big)$: $D$-dependent distribution \\
Host DM & $\log\mathcal{N}(\mu_{\ln}=3.4,\, \sigma_{\ln}=0.5)$ \\
Intrinsic Flux $F$ & Log-Uniform, $p(F) \propto F^{-1}$ \\
\bottomrule
\end{tabular}
\end{table}

The total DM is modeled as the sum of three components: 
\[
\mathrm{DM}_{\rm tot} \;\approx \; \mathrm{DM}_{\rm MW}
\;+\; \mathrm{DM}_{\rm IGM}(z)
\;+\; \frac{\mathrm{DM}_{\rm host}}{1+z},
\]
where $z$ is implied by $D$ via the assumed cosmology. $\mathrm{DM}_{\mathrm{Gal}}$ is the Galactic contribution, $\mathrm{DM}_{\mathrm{IGM}}$ is the contribution from the intergalactic medium (IGM), and $\mathrm{DM}_{\mathrm{host}}$ is the host galaxy contribution. We neglect a separate MW halo term for simplicity.

We utilize the YMW16 model \citep{2017ApJ...835...29Y}, as implemented in the PyGEDM software\footnote{\href{https://github.com/FRBs/pygedm}{https://github.com/FRBs/pygedm}} \citep[][]{2021PASA...38...38P, 2002astro.ph..7156C, 2003astro.ph..1598C, 2003ApJS..146..407F, 2017ApJ...835...29Y, 2020ApJ...888..105Y, 2020ApJ...895L..49P}, to derive the Galactic contribution $\mathrm{DM}_{\mathrm{Gal}}$ for any direction in any pixel. We assume a 20\% uncertainty on the Galactic DM.

The IGM contribution is derived from its empirical dependence on distance $D$ \citep[see e.g.,][]{2003ApJ...598L..79I}, with the uncertainty propagated from the distance uncertainty. In our current model we do not include additional scatter in ${\rm DM}_{\rm IGM}$ from potential intervening galaxies or galaxy clusters along the line of sight. 
While such structures can introduce significant sightline-to-sightline variance at higher redshifts \citep[e.g.,][]{2020Natur.581..391M}, the inferred distance of our event is small on cosmological scales, so this variance is expected to be minor. 
We therefore assume the contribution is negligible for this work, but it can be incorporated in future analyses if required. For the host galaxy contribution, we assume a log-normal distribution with parameters described in the table. 

Finally, we adopt a uniform prior for the signal's arrival time and a log-uniform prior for its intrinsic flux.

\subsection{Analysis Workflow}
Directly applying this framework to the full dedispersed dataset is computationally intractable. We therefore adopt a natural two-stage analysis approach.

In the first stage, our search pipeline (see \autoref{sec:dedispersion}, \autoref{sec:cand}) performs a broad search over the data for candidate events. If a candidate is successfully detected, we proceed to the second stage, where we analyze the event using our Bayesian framework. In cases where multiple high-SNR candidates are identified, we perform a mixture-model analysis using the likelihood model described in the previous subsection. Specifically, the Bayesian evidence quantifies how well the full physical model explains the data for each candidate. Posterior probabilities can then be compared via their respective evidence values, allowing us to identify the most plausible counterpart. The final parameter estimates are derived from the posterior distributions produced by the sampler, using a mixture model if multiple candidates are comparably likely.

In the case of a non-detection, we can place upper limits by performing inference using a single null candidate and marginalize over all model parameters to obtain the posterior distributions. We then report the 95\% posterior quantile as the upper limit. Using the inferred posterior on flux and source distance, we can also compute the posterior distribution of the isotropic-equivalent radio luminosity.

\subsection{Source Constraints from Nested Sampling}

We performed Bayesian inference for both detection (high-SNR injected signal) and non-detection scenarios using nested sampling with the \texttt{dynesty} sampler \citep[][]{2020MNRAS.493.3132S}. The nested sampling was run with 500 live points using the \texttt{rwalk} sampling strategy.

For the detection case, we use the same injected candidate described in \autoref{sec:dedispersion}. From the posterior samples, we derive physical parameters, which are presented in \autoref{fig:detection_vs_nondetection}.

\begin{figure}[h!]
    \centering
    \includegraphics[width=0.5\textwidth]{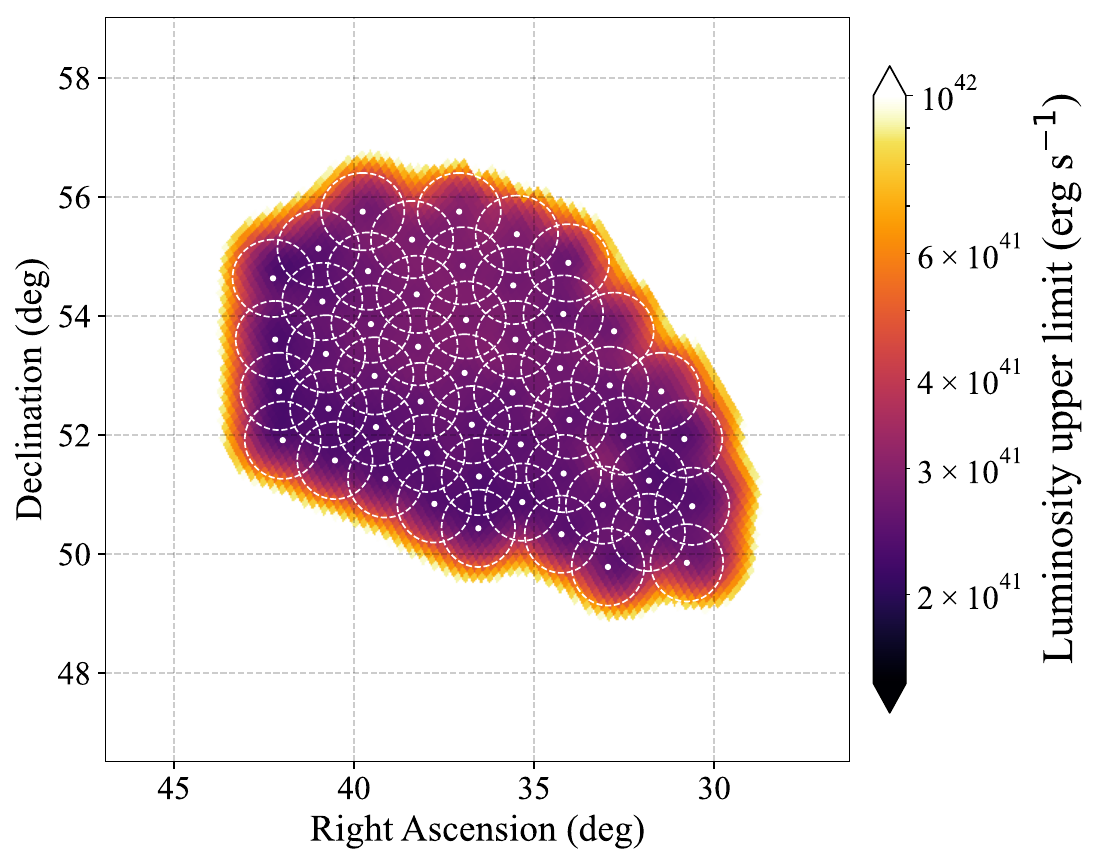}
    \caption{The 95\% confidence luminosity upper limit. The dashed white circles show the full width at half maximum (FWHM) of the individual telescope beams at the central frequency of the used band tiled across the region.}
    \label{fig:limit_map}
\end{figure}

For the non-detection case, we also show posterior distributions in the same figure. The posterior on luminosity $L$ was constructed by combining flux upper limits with the distance prior and is presented in \autoref{fig:limit_map}. We report the band-integrated luminosity
\[
  L \;=\; 4\pi D^2\,F\,\Delta\nu,
\]
where $D$ is the luminosity distance and $\Delta\nu$ is the
bandwidth used in the search.
We assume a flat $F$ across $\Delta\nu$.
The 95\,\% credible upper limit for luminosity calculated exclusively over the region covered by our beams turns out to be:
\[
  L_{95} \;=\;4.0\times10^{41}\;\mathrm{erg\,s^{-1}}.
\]
For the non-detection case, the radio data carry very little additional
information about the sky position, distance, or DM beyond
the GW-informed priors and our adopted DM model. As a result, the posteriors
for $\mathrm{RA}$, $\mathrm{Dec}$, $D$, and the DM components are essentially
identical to those in the LVK skymap and DM priors: the radio non-detection neither selects a single preferred sky location nor
significantly narrows the GW-inferred distance or DM distributions. The only intrinsically new constraint provided by the OVRO-LWA data is therefore this upper limit on the isotropic-equivalent luminosity.

\section{Discussion}
\label{sec:discussion}
\subsection{Comparison with Other Low-Frequency Searches}

We note that our $7\sigma$ fluence threshold of about $150$\,Jy\,ms is comparable to the tightest targeted limits in the literature. In rapid-response MWA searches, \citet{2022PASA...39....3T} reached 80–100\,Jy\,ms for the short GRB 190627A, while a high-time-resolution MWA voltage search on the long GRB 210419A found limits of 77–224\,Jy\,ms over 0.5–10 ms pulses \citep{2022MNRAS.514.2756T}. At 144 MHz, \citet{2024MNRAS.534.2592R} report a candidate coherent flash following GRB 201006A with an observed fluence of $245\pm135$\,Jy\,ms. Our non-detection for \texttt{S250206dm} yields similar limits while targeting a neutron-star–containing GW event rather than a GRB.

Our limit also complements constraints from wide-field, higher-frequency (400--800\,MHz) serendipitous surveys such as CHIME/FRB \citep{2023ApJ...954..154C, 2024ApJ...972..125C}. While CHIME's optimal serendipitous flux-density limits can be as deep as $\lesssim$1--10\,Jy (corresponding to $\lesssim$10--100\,Jy\,ms for a 10\,ms burst), their at-the-time-of-burst flux constraints are typically in the range of a few kJy. This corresponds to a much higher fluence, making our OVRO–LWA limit $\sim$1–2 orders of magnitude more constraining than typical prompt CHIME constraints at the time of the burst.

\subsection{Physical Implications of the Non-Detection}

To place our non-detection in a physical context, we compare our observational upper limit to theoretical predictions. The landscape of proposed coherent emission mechanisms is vast and complex. Below, we evaluate our limit against several representative models, including semi-analytic estimates and recent numerical simulations. These models, while simplified, provide valuable order-of-magnitude benchmarks for some scenarios, including pre-merger magnetospheric interactions, jet–ISM shocks, and post-merger magnetar remnants.

More detailed constraints—linking low-frequency radiative transfer, plasma conditions, merger ejecta, and the full range of proposed coherent-emission mechanisms—are beyond the scope of this work and could be explored in future studies.

\subsubsection{Pre-Merger Magnetospheric Alignment Model}

We consider the pre-merger magnetospheric interaction mechanism originally proposed by \citet{1996A&A...312..937L} and later developed by \citet{2001MNRAS.322..695H,2013ApJ...768...63L}. In this scenario, the neutron stars are spun up to millisecond periods in the final seconds before coalescence, reviving pulsar-like coherent emission powered by magnetic dipole and quadrupole radiation. The peak luminosity is expected near the moment of first contact of the stellar surfaces \citep{2013ApJ...768...63L}. The corresponding observable flux density is given by \citet[][]{2019MNRAS.489.3316R}:

\begin{equation*}
F_\nu \;\approx\;
2\times10^{8}\,
(1+z)\,
B_{15}^2\,M_{1.4}^3\,R_6^{-1}\,
\nu_{9,\mathrm{obs}}^{-1}\,
D^{-2}\,
\epsilon_r
\;\; \mathrm{Jy},
\label{eq:premerger_flux}
\end{equation*}
where $B_{15}$ is the pre-merger magnetic field in units of $10^{15}\,$G, $M_{1.4}$ is the neutron star mass in 1.4\,M$_\odot$, $R_6$ is the radius in $10^6$\,cm, $\nu_{9,\rm obs}=0.077$ for our observing frequency of $\sim77$\,MHz,  $D$ is the luminosity distance in Gpc, and $\epsilon_r$ is the radio efficiency (typically $\epsilon_r \sim 10^{-4}$ for pulsars).

Adopting the fiducial parameters used in \citet{2019MNRAS.489.3316R}, namely $M_{1.4}=1$, $R_6=1$, $B_{15}=10^{-3}$, and $\epsilon_r=10^{-4}$, and assuming the gravitational-wave distance of $D=0.373$\,Gpc for \texttt{S250206dm}, we find that the predicted flux density is far below the OVRO-LWA detection threshold. However, we note that for much nearer GW events ($D \lesssim 40$\,Mpc) the same model would yield flux densities potentially detectable with the OVRO-LWA.

\subsubsection{Coherent Shock Emission from Jet--ISM Interaction}
The second model we consider is the jet--ISM interaction proposed by \citet{2000A&A...364..655U}, in which a Poynting-flux dominated GRB outflow generates low-frequency electromagnetic waves at the external shock. Following \citet{2019MNRAS.489.3316R} and \citet{2022MNRAS.514.2756T}, the characteristic spectral peak frequency is
\begin{equation*}
\nu_{\rm max} \sim 10^{6}\,(1+z)^{-1/2}\,\epsilon_{B}^{1/2}\ \mathrm{Hz},
\end{equation*}
and for observing frequencies above this peak, the expected flux density is
\begin{equation*}
\begin{aligned}
F_\nu \;\approx\;&\;
0.1\,\epsilon_{B}\,(\beta - 1)\,
\mathcal{F}_\gamma\,
\tau^{-1}\,
\nu_{\rm max}^{-1} \\
&\times
\left(\nu_{\rm obs} / \nu_{\rm max}\right)^{-\beta}
\ \mathrm{erg\,cm^{-2}\,s^{-1}\,Hz^{-1}}.
\end{aligned}
\end{equation*}
Here $\epsilon_{B}$ is the fraction of magnetic energy in the relativistic jet, $\mathcal{F}_\gamma$ is the $\gamma$-ray fluence of the associated GRB, $\tau$ is the intrinsic pulse duration, $\beta$ is the spectral index of the coherent emission, and $\nu_{\rm obs}$ and $\nu_{\rm max}$ denote the observing and peak frequencies, respectively. We adopt the following values for $\beta \simeq 1.6$ and $\epsilon_{B} \sim 10^{-3}$. All quantities are expressed in cgs units.

We do not have a measured prompt $\gamma$-ray fluence associated with \texttt{S250206dm}.  
To estimate it, we assume a typical isotropic-equivalent $\gamma$-ray energy
$E_\gamma \sim 10^{50}\,$erg at a distance $D = 373\,$Mpc, which
corresponds to a prompt $\gamma$-ray fluence
\begin{equation*}
\mathcal{F}_\gamma \;\simeq\;
E_\gamma\,(4\pi D^{2})^{-1}
\;\approx\; 6\times 10^{-6}\ \mathrm{erg\,cm^{-2}}.
\end{equation*}
We obtain the corresponding radio flux density at 77\,MHz, assuming a
$1$\,ms coherent pulse, of 
$F_\nu \sim 10^{6}\ \mathrm{Jy}$,
which lies many orders of magnitude above our OVRO-LWA upper limits.

However, the simple \citet{2000A&A...364..655U} coherent–shock model does not account for several propagation effects that can strongly suppress or prevent MHz emission from escaping, including induced Compton (and Raman) scattering \citep[e.g.,][]{2007ApJ...658L...1M,2008ApJ...682.1443L}, free–free absorption in dense local environments \citep[e.g.,][]{2004MNRAS.348..999I},  absorption by merger ejecta in compact-binary events \citep[e.g.,][]{2018PASJ...70...39Y}.

\subsubsection{Pulsar-Like Emission from a Post-Merger Magnetar}

A third class of models invokes pulsar-like, dipole spin-down emission from a long-lived merger remnant. In the scenario proposed by \citet{2013PASJ...65L..12T}, the merger of two neutron stars produces a rapidly rotating, highly magnetized neutron star. Magnetic braking then powers coherent radio emission in a way analogous to ordinary radio pulsars, potentially generating FRB-like bursts for as long as the magnetar remains stable. The expected flux density can be written as \citep[e.g.,][]{2019MNRAS.489.3316R,2023ApJ...954..154C}
\begin{equation*}
F_\nu \;\approx\;
8\times 10^{7}\,
\nu_{\rm obs}^{-1}\,
\epsilon_{r}\,
D^{-2}\,
B_{15}^{2}\,
R_{6}^{6}\,
P_{-3}^{-4} 
\ \mathrm{Jy},
\end{equation*}
where $\nu_{\rm obs}$ is the observing frequency in MHz, $\epsilon_{r}$ is the efficiency for converting spin-down power into coherent radio emission, $D$ is the luminosity distance in units of 1\,Gpc, $B_{15}$ is the surface dipole field in units of $10^{15}\,$G, $R_{6}$ is the stellar radius in units of $10^{6}\,$cm, and $P_{-3}$ is the spin period in units of $10^{-3}\,$s. 

As a fiducial example, we consider a millisecond magnetar with
$B_{15} = 1$, $R_{6} = 1$, $P_{-3} = 1$, and a pulsar-like radio efficiency
$\epsilon_{r} \sim 10^{-4}$. For \texttt{S250206dm}, adopting the same luminosity distance
of $373\,$Mpc and the OVRO-LWA central
frequency $\nu_{\rm obs} \approx 77\,$MHz, we obtain
$F_\nu \sim 7\times 10^{2}\ \mathrm{Jy},$ above our OVRO-LWA upper limits.

In practice, this pulsar-like post-merger emission model is highly uncertain for
\texttt{S250206dm}: it requires that the merger remnant be a stable (or at least long-lived) magnetar rather than a promptly collapsing black hole, and the
coherent emission is expected to be strongly beamed along the magnetar rotation axis. Moreover, the MHz emission may be significantly attenuated or completely
blocked by the dynamical ejecta and surrounding plasma (see discussion above), so this estimate should be regarded as an optimistic benchmark rather than a robust prediction.

\subsubsection{Magnetar Collapse (Blitzar-Like) Model}

If the post-merger remnant is a supramassive magnetar, its eventual collapse to a black hole can drive a short burst of coherent radio emission via magnetic reconnection in the magnetosphere \citep{2014A&A...562A.137F,2014ApJ...780L..21Z}. Following the formulation of \citet{2019MNRAS.489.3316R}, the observed flux density for small redshifts can be written as
\begin{equation*}
\begin{aligned}
F_\nu \;\approx\;&\;
(4\pi D^{2}\,
\tau)^{-1}\,\epsilon\,
E_{B}\,|\alpha+1|\,
\\[4pt]
&\times\nu_{p}^{-(\alpha+1)}\,\nu_{\rm obs}^{\alpha}\, \mathrm{erg\,cm^{-2}\,s^{-1}\,Hz^{-1}},
\end{aligned}
\end{equation*}
where $\epsilon$ is the efficiency for converting magnetic energy into coherent radio emission, $E_{B}$ is the magnetic energy released during the collapse, $D$ is the luminosity distance, $\tau$ is the intrinsic burst duration, $\alpha$ is the spectral index of the coherent emission, $\nu_{p}$ is the plasma frequency, $\nu_{\rm obs}$ is the observing frequency. All quantities are expressed in cgs units.

Adopting the fiducial parameters used by \citet{2014ApJ...780L..21Z,2019MNRAS.489.3316R},
$E_B \sim 1.7\times 10^{47}\,\mathrm{erg}$,
$\tau \sim 10^{-3}\,\mathrm{s}$,
$\alpha \sim -3$,
efficiency $\epsilon \sim 10^{-4}$,
and a plasma cutoff frequency $\nu_p \sim 10^{3}\,\mathrm{Hz}$,
we evaluate the predicted flux density at the OVRO-LWA observing frequency
and the luminosity distance of
\texttt{S250206dm}.  
For $\nu_p = 1~\mathrm{kHz}$, the resulting flux density is
far below our $\sim 10^{2}\,$Jy sensitivity threshold.  
However, the flux is highly sensitive to the assumed plasma frequency.  
If we instead adopt $\nu_p \sim 10^{6}\,\mathrm{Hz}$ (corresponding to an electron density
$n_e \sim 10^{4}\,\mathrm{cm}^{-3}$), the predicted flux increases, yielding
$F_\nu \sim \mathrm{few}\times10^{2}\ \mathrm{Jy},$ comparable to our observational limits.

\subsubsection{Numerical Simulation Constraints}

Beyond analytic estimates, several recent numerical studies have modeled the electromagnetic output of compact-object mergers using global general-relativistic force-free electrodynamics simulations of the binary magnetosphere. Three-dimensional merger simulations by \citet{2022MNRAS.515.2710M} predict precursor Poynting luminosities of $L_{\rm EM}\sim10^{44}$--$10^{45}\,\mathrm{erg\,s^{-1}}$ for neutron-star magnetic fields $B\sim10^{12}$\,G, implying possible coherent radio powers of $L_{\rm R}\sim10^{40}$--$10^{43}\,\mathrm{erg\,s^{-1}}$ for plausible conversion efficiencies $\epsilon_r\sim10^{-4}$--$10^{-2}$. Black hole–neutron star simulations yield somewhat lower values, $L_{\rm EM}\sim10^{40}$--$10^{42}\,\mathrm{erg\,s^{-1}}$ \citep{2023ApJ...956L..33M}. It is important to note, however, that these models primarily predict emission peaking at high frequencies ($\sim 9$--$20$\,GHz), leaving the low-frequency regime, such as the 70-80 MHz band, as a largely unexplored parameter space for these precursors.

Our OVRO-LWA 95\,\% upper limit on the isotropic-equivalent luminosity, $L_{95} = 4.0\times10^{41}\ \mathrm{erg\,s^{-1}},$
lies at the upper end of the luminosity range suggested by these simulation-based models. Consequently, while our non-detection does not yet rule out most simulated scenarios, it begins to probe the brightest end of the predicted coherent-emission parameter space for a neutron-star–containing merger at the distance of \texttt{S250206dm}.

\subsection{Extending Single-Event Limits to Population Inference}
A single non-detection constrains the luminosity of prompt, coherent radio bursts for that event, but the real power comes from stacking results across many mergers. If $f(L_\star)$ is the fraction of mergers that emit above a luminosity threshold $L_\star$, then after $N$ independent searches with average per-event detectability $\bar p(L_\star)$, the 95\% confidence upper limit on this fraction is given by Poisson statistics:
\[
f_{95}(L_\star) \;\lesssim\; \frac{3}{N\,\bar p(L_\star)} \,.
\]

Here $\bar p(L_\star)$ is the typical chance of detection if such emission occurred, folding in the searched time window, DM coverage, fraction of the GW sky posterior coherently tiled, and within-beam detectability set by beam overlap, gain variation, and distance. For \texttt{S250206dm}, the time and DM coverage were nearly complete, while sky coverage was limited to $\sim$50\% of the localization region. At our sensitivity limit of $L_\star \!\sim\! 4\times10^{41}\,\mathrm{erg\,s^{-1}}$, this scaling implies that roughly ten similar, well-covered events would already exclude the hypothesis that all such mergers emit above this luminosity, while a few tens of events would constrain the emitting fraction to $\lesssim 50\%$.

This framework has direct implications for the current O4 and upcoming O5 runs. Typical BNS events in O4 occur at a few hundred Mpc, while NSBH systems extend farther \citep[e.g.,][]{2023ApJ...958..158K}, reducing their detectability at fixed flux sensitivity since $L \propto D^2$. Thus, the greatest leverage will come from nearby, well-localized BNS-like mergers, especially when most of the GW sky posterior can be coherently tiled. Our present result therefore provides both a concrete luminosity constraint for \texttt{S250206dm} and a first building block toward population-level bounds on prompt, low-frequency radio emission from compact binary mergers.

We further note that an early kilonova identification would localize the source and constrain the distance and host-galaxy DM, allowing us to beamform over a much smaller region. This would speed processing and enable stronger priors in the Bayesian analysis, yielding significantly tighter flux limits.

\section{Summary}
\label{sec:sum}
We carried out a targeted search for prompt, coherent radio emission associated with the gravitational-wave merger \texttt{S250206dm}, leveraging the OVRO-LWA and its dedicated \textit{Time Machine} system. Given the early indication of a neutron-star–containing merger (BNS or NSBH), this was a particularly well-motivated target for prompt coherent emission. The system continuously stores raw voltages in a ring buffer and enables offline beamforming, dedispersion, and candidate searches across a wide field of view. For this event, we analyzed a 30-minute window starting shortly after the merger and preceding the preliminary alert that initiated data saving. We focused on the 69--86\,MHz band, tiling the 50\% localization region with 51 coherent beams, with an additional beam formed toward FRB~20250206A.

Our transient search used GPU-accelerated dedispersion, optimized on both local and cloud infrastructure, and incorporated boxcar filtering over a broad range of timescales (1–200\,ms). No astrophysical candidates were identified above our conservative $7\sigma$ fluence threshold of $\sim$150\,Jy\,ms. We validated our sensitivity using injected signals and pulsar benchmarks.

To interpret the non-detection and to infer physical parameters from future searches and detections, we developed a Bayesian inference framework that incorporates the sky position, distance, dispersion measure components, and OVRO-LWA characteristics, conditioned on the LIGO/Virgo/KAGRA skymap and established DM models. This allowed us to place a 95\% credible upper limit on the source's isotropic-equivalent radio luminosity of $L_{95} = 4.0 \times 10^{41}\,\mathrm{erg\,s^{-1}}$ within the searched region. The same framework is fully general and can be applied to radio datasets from other facilities. Because the forward model depends only on the instrument response, frequency coverage, and fluence sensitivity, limits or detections from instruments such as LOFAR, MWA, CHIME/FRB, or ASKAP can be re-expressed in terms of intrinsic luminosity, distance, and DM components, enabling direct comparisons across instruments.

Our $7\sigma$ fluence threshold of $\sim$150\,Jy\,ms is comparable to the tightest targeted, low-frequency limits for GRBs and is more constraining than typical prompt limits from wide-field surveys. We place this non-detection in the context of several coherent emission models. We find that while pre-merger models are likely undetectable at the distance ($D \sim 370$\,Mpc) of the GW event, our luminosity limit places constraints on jet--ISM shock models and begins to probe the brightest end of the parameter space for post-merger magnetar, blitzar, and recent simulation-based scenarios.

While a single non-detection sets a valuable event-specific constraint, robustly establishing or ruling out an association between compact binary mergers and prompt low-frequency radio bursts requires a larger sample. We find that of order ten well-covered events are needed to exclude the hypothesis that all mergers emit above our current sensitivity threshold. Nonetheless, the present analysis represents an important milestone: it delivers the first millisecond-timescale, low-frequency search for transient emission from a GW event with the OVRO-LWA.

Looking ahead, next-generation wide-field facilities—including DSA-2000, CHORD, and the SKA \citep[e.g.,][]{2019BAAS...51g.255H,2019clrp.2020...28V,2019arXiv191212699B,2015aska.confE.174B}—will offer large improvements in sensitivity, bandwidth, and rapid tiling capability. These systems will be able to probe much fainter coherent bursts at cosmologically relevant distances, greatly increasing the number of mergers for which meaningful limits can be placed and enabling population-level tests of prompt-emission models. Their combination of continuous monitoring, expansive fields of view, and low-latency response will allow real-time searches for both GW events and GRBs, opening the possibility of detecting—or definitively excluding—fast coherent precursors and post-merger flashes across diverse environments. Continued improvements in alert latency, GPU efficiency, bandwidth, and observing strategies will further enhance future search sensitivity, moving us closer to detecting prompt radio counterparts from GW sources.

\begin{acknowledgments}
\section{Acknowledgments}
We thank the anonymous referee and Elias Roland Most for useful comments on the manuscript. This material is based in part upon work supported by the National Science Foundation under grant numbers AST-1828784, AST-1911199, the Simons Foundation (668346, JPG), the Wilf Family Foundation and Mt. Cuba Astronomical Foundation.
\end{acknowledgments}

%

\vspace{5mm}






\bibliography{time_machine_search}{}

@ARTICLE{2024MNRAS.534.2592R,
       author = {{Rowlinson}, A. and {de Ruiter}, I. and {Starling}, R.~L.~C. and {Rajwade}, K.~M. and {Hennessy}, A. and {Wijers}, R.~A.~M.~J. and {Anderson}, G.~E. and {Mevius}, M. and {Ruhe}, D. and {Gourdji}, K. and {van der Horst}, A.~J. and {ter Veen}, S. and {Wiersema}, K.},
        title = "{A candidate coherent radio flash following a neutron star merger}",
      journal = {\mnras},
         year = 2024,
        month = nov,
       volume = {534},
       number = {3},
        pages = {2592-2608},
          doi = {10.1093/mnras/stae2234},
archivePrefix = {arXiv},
       eprint = {2312.04237},
 primaryClass = {astro-ph.HE},
       adsurl = {https://ui.adsabs.harvard.edu/abs/2024MNRAS.534.2592R}
}

@ARTICLE{2025ApJ...985..265K,
       author = {{Kosogorov}, Nikita and {Hallinan}, Gregg and {Law}, Casey and {Hickish}, Jack and {Dowell}, Jayce and {Anderson}, Marin M. and {Bowman}, Judd D. and {Byrne}, Ruby and {Catha}, Morgan and {Chen}, Bin and {Chhabra}, Sherry and {D'Addario}, Larry and {Davis}, Ivey and {Elder}, Katherine and {Gary}, Dale and {Harnach}, Charlie and {Hellbourg}, Greg and {Hobbs}, Rick and {Hodge}, David and {Hodges}, Mark and {Huang}, Yuping and {Isella}, Andrea and {Jacobs}, Daniel C. and {Kemby}, Ghislain and {Klinefelter}, John T. and {Kolopanis}, Matthew and {Lamb}, James and {Mahesh}, Nivedita and {Mondal}, Surajit and {O'Donnell}, Brian and {Plant}, Kathryn and {Posner}, Corey and {Prayag}, Vinand and {Rizo}, Andres and {Romero-Wolf}, Andrew and {Shi}, Jun and {Taylor}, Greg and {Virgin}, Mike and {Vydula}, Akshatha and {Weinreb}, Sandy and {Woody}, David and {Yu}, Sijie and {Zhang}, Peijin and {Zhao}, Yifan (Amy)},
        title = "{Implementing Continuous All-sky Monitoring with the OVRO-LWA to Identify Prompt and Precursor Counterparts of Gravitational Wave Events}",
      journal = {\apj},
         year = 2025,
        month = jun,
       volume = {985},
       number = {2},
          eid = {265},
        pages = {265},
          doi = {10.3847/1538-4357/add014},
       adsurl = {https://ui.adsabs.harvard.edu/abs/2025ApJ...985..265K}
}

@ARTICLE{2017Natur.551...67P,
       author = {{Pian}, E. and {D'Avanzo}, P. and {Benetti}, S. and {Branchesi}, M. and {Brocato}, E. and {Campana}, S. and {Cappellaro}, E. and {Covino}, S. and {D'Elia}, V. and {Fynbo}, J.~P.~U. and {Getman}, F. and {Ghirlanda}, G. and {Ghisellini}, G. and {Grado}, A. and {Greco}, G. and {Hjorth}, J. and {Kouveliotou}, C. and {Levan}, A. and {Limatola}, L. and {Malesani}, D. and {Mazzali}, P.~A. and {Melandri}, A. and {M{\o}ller}, P. and {Nicastro}, L. and {Palazzi}, E. and {Piranomonte}, S. and {Rossi}, A. and {Salafia}, O.~S. and {Selsing}, J. and {Stratta}, G. and {Tanaka}, M. and {Tanvir}, N.~R. and {Tomasella}, L. and {Watson}, D. and {Yang}, S. and {Amati}, L. and {Antonelli}, L.~A. and {Ascenzi}, S. and {Bernardini}, M.~G. and {Bo{\"e}r}, M. and {Bufano}, F. and {Bulgarelli}, A. and {Capaccioli}, M. and {Casella}, P. and {Castro-Tirado}, A.~J. and {Chassande-Mottin}, E. and {Ciolfi}, R. and {Copperwheat}, C.~M. and {Dadina}, M. and {De Cesare}, G. and {di Paola}, A. and {Fan}, Y.~Z. and {Gendre}, B. and {Giuffrida}, G. and {Giunta}, A. and {Hunt}, L.~K. and {Israel}, G.~L. and {Jin}, Z. -P. and {Kasliwal}, M.~M. and {Klose}, S. and {Lisi}, M. and {Longo}, F. and {Maiorano}, E. and {Mapelli}, M. and {Masetti}, N. and {Nava}, L. and {Patricelli}, B. and {Perley}, D. and {Pescalli}, A. and {Piran}, T. and {Possenti}, A. and {Pulone}, L. and {Razzano}, M. and {Salvaterra}, R. and {Schipani}, P. and {Spera}, M. and {Stamerra}, A. and {Stella}, L. and {Tagliaferri}, G. and {Testa}, V. and {Troja}, E. and {Turatto}, M. and {Vergani}, S.~D. and {Vergani}, D.},
        title = "{Spectroscopic identification of r-process nucleosynthesis in a double neutron-star merger}",
      journal = {\nat},
         year = 2017,
        month = nov,
       volume = {551},
       number = {7678},
        pages = {67-70},
          doi = {10.1038/nature24298},
archivePrefix = {arXiv},
       eprint = {1710.05858},
 primaryClass = {astro-ph.HE},
       adsurl = {https://ui.adsabs.harvard.edu/abs/2017Natur.551...67P}
}

@ARTICLE{2017Sci...358.1579H,
       author = {{Hallinan}, G. and {Corsi}, A. and {Mooley}, K.~P. and {Hotokezaka}, K. and {Nakar}, E. and {Kasliwal}, M.~M. and {Kaplan}, D.~L. and {Frail}, D.~A. and {Myers}, S.~T. and {Murphy}, T. and {De}, K. and {Dobie}, D. and {Allison}, J.~R. and {Bannister}, K.~W. and {Bhalerao}, V. and {Chandra}, P. and {Clarke}, T.~E. and {Giacintucci}, S. and {Ho}, A.~Y.~Q. and {Horesh}, A. and {Kassim}, N.~E. and {Kulkarni}, S.~R. and {Lenc}, E. and {Lockman}, F.~J. and {Lynch}, C. and {Nichols}, D. and {Nissanke}, S. and {Palliyaguru}, N. and {Peters}, W.~M. and {Piran}, T. and {Rana}, J. and {Sadler}, E.~M. and {Singer}, L.~P.},
        title = "{A radio counterpart to a neutron star merger}",
      journal = {Science},
         year = 2017,
        month = dec,
       volume = {358},
       number = {6370},
        pages = {1579-1583},
          doi = {10.1126/science.aap9855},
archivePrefix = {arXiv},
       eprint = {1710.05435},
 primaryClass = {astro-ph.HE},
       adsurl = {https://ui.adsabs.harvard.edu/abs/2017Sci...358.1579H}
}

@ARTICLE{2018Natur.561..355M,
       author = {{Mooley}, K.~P. and {Deller}, A.~T. and {Gottlieb}, O. and {Nakar}, E. and {Hallinan}, G. and {Bourke}, S. and {Frail}, D.~A. and {Horesh}, A. and {Corsi}, A. and {Hotokezaka}, K.},
        title = "{Superluminal motion of a relativistic jet in the neutron-star merger GW170817}",
      journal = {\nat},
         year = 2018,
        month = sep,
       volume = {561},
       number = {7723},
        pages = {355-359},
          doi = {10.1038/s41586-018-0486-3},
archivePrefix = {arXiv},
       eprint = {1806.09693},
 primaryClass = {astro-ph.HE},
       adsurl = {https://ui.adsabs.harvard.edu/abs/2018Natur.561..355M}
}

@ARTICLE{2017Natur.551...85A,
       author = {{Abbott}, B.~P. and {Abbott}, R. and {Abbott}, T.~D. and {Acernese}, F. and {Ackley}, K. and {Adams}, C. and {Adams}, T. and {Addesso}, P. and {Adhikari}, R.~X. and {Adya}, V.~B. and {Affeldt}, C. and {Afrough}, M. and {Agarwal}, B. and {Agathos}, M. and {Agatsuma}, K. and {Aggarwal}, N. and {Aguiar}, O.~D. and {Aiello}, L. and {Ain}, A. and {Ajith}, P. and {Allen}, B. and {Allen}, G. and {Allocca}, A. and {Altin}, P.~A. and {Amato}, A. and {Ananyeva}, A. and {Anderson}, S.~B. and {Anderson}, W.~G. and {Angelova}, S.~V. and {Antier}, S. and {Appert}, S. and {Arai}, K. and {Araya}, M.~C. and {Areeda}, J.~S. and {Arnaud}, N. and {Arun}, K.~G. and {Ascenzi}, S. and {Ashton}, G. and {Ast}, M. and {Aston}, S.~M. and {Astone}, P. and {Atallah}, D.~V. and {Aufmuth}, P. and {Aulbert}, C. and {Aultoneal}, K. and {Austin}, C. and {Avila-Alvarez}, A. and {Babak}, S. and {Bacon}, P. and {Bader}, M.~K.~M. and {Bae}, S. and {Baker}, P.~T. and {Baldaccini}, F. and {Ballardin}, G. and {Ballmer}, S.~W. and {Banagiri}, S. and {Barayoga}, J.~C. and {Barclay}, S.~E. and {Barish}, B.~C. and {Barker}, D. and {Barkett}, K. and {Barone}, F. and {Barr}, B. and {Barsotti}, L. and {Barsuglia}, M. and {Barta}, D. and {Bartlett}, J. and {Bartos}, I. and {Bassiri}, R. and {Basti}, A. and {Batch}, J.~C. and {Bawaj}, M. and {Bayley}, J.~C. and {Bazzan}, M. and {B{\'e}csy}, B. and {Beer}, C. and {Bejger}, M. and {Belahcene}, I. and {Bell}, A.~S. and {Berger}, B.~K. and {Bergmann}, G. and {Bero}, J.~J. and {Berry}, C.~P.~L. and {Bersanetti}, D. and {Bertolini}, A. and {Betzwieser}, J. and {Bhagwat}, S. and {Bhandare}, R. and {Bilenko}, I.~A. and {Billingsley}, G. and {Billman}, C.~R. and {Birch}, J. and {Birney}, R. and {Birnholtz}, O. and {Biscans}, S. and {Biscoveanu}, S. and {Bisht}, A. and {Bitossi}, M. and {Biwer}, C. and {Bizouard}, M.~A. and {Blackburn}, J.~K. and {Blackman}, J. and {Blair}, C.~D. and {Blair}, D.~G. and {Blair}, R.~M. and {Bloemen}, S. and {Bock}, O. and {Bode}, N. and {Boer}, M. and {Bogaert}, G. and {Bohe}, A. and {Bondu}, F. and {Bonilla}, E. and {Bonnand}, R. and {Boom}, B.~A. and {Bork}, R. and {Boschi}, V. and {Bose}, S. and {Bossie}, K. and {Bouffanais}, Y. and {Bozzi}, A. and {Bradaschia}, C. and {Brady}, P.~R. and {Branchesi}, M. and {Brau}, J.~E. and {Briant}, T. and {Brillet}, A. and {Brinkmann}, M. and {Brisson}, V. and {Brockill}, P. and {Broida}, J.~E. and {Brooks}, A.~F. and {Brown}, D.~A. and {Brown}, D.~D. and {Brunett}, S. and {Buchanan}, C.~C. and {Buikema}, A. and {Bulik}, T. and {Bulten}, H.~J. and {Buonanno}, A. and {Buskulic}, D. and {Buy}, C. and {Byer}, R.~L. and {Cabero}, M. and {Cadonati}, L. and {Cagnoli}, G. and {Cahillane}, C. and {Bustillo}, J. Calder{\'o}n and {Callister}, T.~A. and {Calloni}, E. and {Camp}, J.~B. and {Canepa}, M. and {Canizares}, P. and {Cannon}, K.~C. and {Cao}, H. and {Cao}, J. and {Capano}, C.~D. and {Capocasa}, E. and {Carbognani}, F. and {Caride}, S. and {Carney}, M.~F. and {Diaz}, J. Casanueva and {Casentini}, C. and {Caudill}, S. and {Cavagli{\`a}}, M. and {Cavalier}, F. and {Cavalieri}, R. and {Cella}, G. and {Cepeda}, C.~B. and {Cerd{\'a}-Dur{\'a}n}, P. and {Cerretani}, G. and {Cesarini}, E. and {Chamberlin}, S.~J. and {Chan}, M. and {Chao}, S. and {Charlton}, P. and {Chase}, E. and {Chassande-Mottin}, E. and {Chatterjee}, D. and {Chatziioannou}, K. and {Cheeseboro}, B.~D. and {Chen}, H.~Y. and {Chen}, X. and {Chen}, Y. and {Cheng}, H. -P. and {Chia}, H. and {Chincarini}, A. and {Chiummo}, A. and {Chmiel}, T. and {Cho}, H.~S. and {Cho}, M. and {Chow}, J.~H. and {Christensen}, N. and {Chu}, Q. and {Chua}, A.~J.~K. and {Chua}, S. and {Chung}, A.~K.~W. and {Chung}, S. and {Ciani}, G. and {Ciolfi}, R.},
        title = "{A gravitational-wave standard siren measurement of the Hubble constant}",
      journal = {\nat},
         year = 2017,
        month = nov,
       volume = {551},
       number = {7678},
        pages = {85-88},
          doi = {10.1038/nature24471},
archivePrefix = {arXiv},
       eprint = {1710.05835},
 primaryClass = {astro-ph.CO},
       adsurl = {https://ui.adsabs.harvard.edu/abs/2017Natur.551...85A}
}

@ARTICLE{2017Sci...358.1559K,
       author = {{Kasliwal}, M.~M. and {Nakar}, E. and {Singer}, L.~P. and {Kaplan}, D.~L. and {Cook}, D.~O. and {Van Sistine}, A. and {Lau}, R.~M. and {Fremling}, C. and {Gottlieb}, O. and {Jencson}, J.~E. and {Adams}, S.~M. and {Feindt}, U. and {Hotokezaka}, K. and {Ghosh}, S. and {Perley}, D.~A. and {Yu}, P. -C. and {Piran}, T. and {Allison}, J.~R. and {Anupama}, G.~C. and {Balasubramanian}, A. and {Bannister}, K.~W. and {Bally}, J. and {Barnes}, J. and {Barway}, S. and {Bellm}, E. and {Bhalerao}, V. and {Bhattacharya}, D. and {Blagorodnova}, N. and {Bloom}, J.~S. and {Brady}, P.~R. and {Cannella}, C. and {Chatterjee}, D. and {Cenko}, S.~B. and {Cobb}, B.~E. and {Copperwheat}, C. and {Corsi}, A. and {De}, K. and {Dobie}, D. and {Emery}, S.~W.~K. and {Evans}, P.~A. and {Fox}, O.~D. and {Frail}, D.~A. and {Frohmaier}, C. and {Goobar}, A. and {Hallinan}, G. and {Harrison}, F. and {Helou}, G. and {Hinderer}, T. and {Ho}, A.~Y.~Q. and {Horesh}, A. and {Ip}, W. -H. and {Itoh}, R. and {Kasen}, D. and {Kim}, H. and {Kuin}, N.~P.~M. and {Kupfer}, T. and {Lynch}, C. and {Madsen}, K. and {Mazzali}, P.~A. and {Miller}, A.~A. and {Mooley}, K. and {Murphy}, T. and {Ngeow}, C. -C. and {Nichols}, D. and {Nissanke}, S. and {Nugent}, P. and {Ofek}, E.~O. and {Qi}, H. and {Quimby}, R.~M. and {Rosswog}, S. and {Rusu}, F. and {Sadler}, E.~M. and {Schmidt}, P. and {Sollerman}, J. and {Steele}, I. and {Williamson}, A.~R. and {Xu}, Y. and {Yan}, L. and {Yatsu}, Y. and {Zhang}, C. and {Zhao}, W.},
        title = "{Illuminating gravitational waves: A concordant picture of photons from a neutron star merger}",
      journal = {Science},
         year = 2017,
        month = dec,
       volume = {358},
       number = {6370},
        pages = {1559-1565},
          doi = {10.1126/science.aap9455},
archivePrefix = {arXiv},
       eprint = {1710.05436},
 primaryClass = {astro-ph.HE},
       adsurl = {https://ui.adsabs.harvard.edu/abs/2017Sci...358.1559K}
}

@ARTICLE{2004NIMPA.517..154A,
       author = {{Abbott}, B. and {Abbott}, R. and {Adhikari}, R. and {Ageev}, A. and {Allen}, B. and {Amin}, R. and {Anderson}, S.~B. and {Anderson}, W.~G. and {Araya}, M. and {Armandula}, H. and {Asiri}, F. and {Aufmuth}, P. and {Aulbert}, C. and {Babak}, S. and {Balasubramanian}, R. and {Ballmer}, S. and {Barish}, B.~C. and {Barker}, D. and {Barker-Patton}, C. and {Barnes}, M. and {Barr}, B. and {Barton}, M.~A. and {Bayer}, K. and {Beausoleil}, R. and {Belczynski}, K. and {Bennett}, R. and {Berukoff}, S.~J. and {Betzwieser}, J. and {Bhawal}, B. and {Bilenko}, I.~A. and {Billingsley}, G. and {Black}, E. and {Blackburn}, K. and {Bland-Weaver}, B. and {Bochner}, B. and {Bogue}, L. and {Bork}, R. and {Bose}, S. and {Brady}, P.~R. and {Braginsky}, V.~B. and {Brau}, J.~E. and {Brown}, D.~A. and {Brozek}, S. and {Bullington}, A. and {Buonanno}, A. and {Burgess}, R. and {Busby}, D. and {Butler}, W.~E. and {Byer}, R.~L. and {Cadonati}, L. and {Cagnoli}, G. and {Camp}, J.~B. and {Cantley}, C.~A. and {Cardenas}, L. and {Carter}, K. and {Casey}, M.~M. and {Castiglione}, J. and {Chandler}, A. and {Chapsky}, J. and {Charlton}, P. and {Chatterji}, S. and {Chen}, Y. and {Chickarmane}, V. and {Chin}, D. and {Christensen}, N. and {Churches}, D. and {Colacino}, C. and {Coldwell}, R. and {Coles}, M. and {Cook}, D. and {Corbitt}, T. and {Coyne}, D. and {Creighton}, J.~D.~E. and {Creighton}, T.~D. and {Crooks}, D.~R.~M. and {Csatorday}, P. and {Cusack}, B.~J. and {Cutler}, C. and {D'Ambrosio}, E. and {Danzmann}, K. and {Davies}, R. and {Daw}, E. and {DeBra}, D. and {Delker}, T. and {DeSalvo}, R. and {Dhurandar}, S. and {D{\'\i}az}, M. and {Ding}, H. and {Drever}, R.~W.~P. and {Dupuis}, R.~J. and {Ebeling}, C. and {Edlund}, J. and {Ehrens}, P. and {Elliffe}, E.~J. and {Etzel}, T. and {Evans}, M. and {Evans}, T. and {Fallnich}, C. and {Farnham}, D. and {Fejer}, M.~M. and {Fine}, M. and {Finn}, L.~S. and {Flanagan}, {\'E}. and {Freise}, A. and {Frey}, R. and {Fritschel}, P. and {Frolov}, V. and {Fyffe}, M. and {Ganezer}, K.~S. and {Giaime}, J.~A. and {Gillespie}, A. and {Goda}, K. and {Gonz{\'a}lez}, G. and {Go{\ss}ler}, S. and {Grandcl{\'e}ment}, P. and {Grant}, A. and {Gray}, C. and {Gretarsson}, A.~M. and {Grimmett}, D. and {Grote}, H. and {Grunewald}, S. and {Guenther}, M. and {Gustafson}, E. and {Gustafson}, R. and {Hamilton}, W.~O. and {Hammond}, M. and {Hanson}, J. and {Hardham}, C. and {Harry}, G. and {Hartunian}, A. and {Heefner}, J. and {Hefetz}, Y. and {Heinzel}, G. and {Heng}, I.~S. and {Hennessy}, M. and {Hepler}, N. and {Heptonstall}, A. and {Heurs}, M. and {Hewitson}, M. and {Hindman}, N. and {Hoang}, P. and {Hough}, J. and {Hrynevych}, M. and {Hua}, W. and {Ingley}, R. and {Ito}, M. and {Itoh}, Y. and {Ivanov}, A. and {Jennrich}, O. and {Johnson}, W.~W. and {Johnston}, W. and {Jones}, L. and {Jungwirth}, D. and {Kalogera}, V. and {Katsavounidis}, E. and {Kawabe}, K. and {Kawamura}, S. and {Kells}, W. and {Kern}, J. and {Khan}, A. and {Killbourn}, S. and {Killow}, C.~J. and {Kim}, C. and {King}, C. and {King}, P. and {Klimenko}, S. and {Kloevekorn}, P. and {Koranda}, S. and {K{\"o}tter}, K. and {Kovalik}, J. and {Kozak}, D. and {Krishnan}, B. and {Landry}, M. and {Langdale}, J. and {Lantz}, B. and {Lawrence}, R. and {Lazzarini}, A. and {Lei}, M. and {Leonhardt}, V. and {Leonor}, I. and {Libbrecht}, K. and {Lindquist}, P. and {Liu}, S. and {Logan}, J. and {Lormand}, M. and {Lubinski}, M. and {L{\"u}ck}, H. and {Lyons}, T.~T. and {Machenschalk}, B. and {MacInnis}, M. and {Mageswaran}, M. and {Mailand}, K. and {Majid}, W. and {Malec}, M. and {Mann}, F. and {Marin}, A. and {M{\'a}rka}, S. and {Maros}, E. and {Mason}, J. and {Mason}, K.},
        title = "{Detector description and performance for the first coincidence observations between LIGO and GEO}",
      journal = {Nuclear Instruments and Methods in Physics Research A},
         year = 2004,
        month = jan,
       volume = {517},
       number = {1-3},
        pages = {154-179},
          doi = {10.1016/j.nima.2003.11.124},
archivePrefix = {arXiv},
       eprint = {gr-qc/0308043},
 primaryClass = {gr-qc},
       adsurl = {https://ui.adsabs.harvard.edu/abs/2004NIMPA.517..154A}
}

@ARTICLE{2012JInst...7.3012A,
       author = {{Accadia}, T. and {Acernese}, F. and {Alshourbagy}, M. and {Amico}, P. and {Antonucci}, F. and {Aoudia}, S. and {Arnaud}, N. and {Arnault}, C. and {Arun}, K.~G. and {Astone}, P. and {Avino}, S. and {Babusci}, D. and {Ballardin}, G. and {Barone}, F. and {Barrand}, G. and {Barsotti}, L. and {Barsuglia}, M. and {Basti}, A. and {Bauer}, Th S. and {Beauville}, F. and {Bebronne}, M. and {Bejger}, M. and {Beker}, M.~G. and {Bellachia}, F. and {Belletoile}, A. and {Beney}, J.~L. and {Bernardini}, M. and {Bigotta}, S. and {Bilhaut}, R. and {Birindelli}, S. and {Bitossi}, M. and {Bizouard}, M.~A. and {Blom}, M. and {Boccara}, C. and {Boget}, D. and {Bondu}, F. and {Bonelli}, L. and {Bonnand}, R. and {Boschi}, V. and {Bosi}, L. and {Bouedo}, T. and {Bouhou}, B. and {Bozzi}, A. and {Bracci}, L. and {Braccini}, S. and {Bradaschia}, C. and {Branchesi}, M. and {Briant}, T. and {Brillet}, A. and {Brisson}, V. and {Brocco}, L. and {Bulik}, T. and {Bulten}, H.~J. and {Buskulic}, D. and {Buy}, C. and {Cagnoli}, G. and {Calamai}, G. and {Calloni}, E. and {Campagna}, E. and {Canuel}, B. and {Carbognani}, F. and {Carbone}, L. and {Cavalier}, F. and {Cavalieri}, R. and {Cecchi}, R. and {Cella}, G. and {Cesarini}, E. and {Chassande-Mottin}, E. and {Chatterji}, S. and {Chiche}, R. and {Chincarini}, A. and {Chiummo}, A. and {Christensen}, N. and {Clapson}, A.~C. and {Cleva}, F. and {Coccia}, E. and {Cohadon}, P. -F. and {Colacino}, C.~N. and {Colas}, J. and {Colla}, A. and {Colombini}, M. and {Conforto}, G. and {Corsi}, A. and {Cortese}, S. and {Cottone}, F. and {Coulon}, J. -P. and {Cuoco}, E. and {D'Antonio}, S. and {Daguin}, G. and {Dari}, A. and {Dattilo}, V. and {David}, P.~Y. and {Davier}, M. and {Day}, R. and {Debreczeni}, G. and {De Carolis}, G. and {Dehamme}, M. and {Del Fabbro}, R. and {Del Pozzo}, W. and {del Prete}, M. and {Derome}, L. and {De Rosa}, R. and {DeSalvo}, R. and {Dialinas}, M. and {Di Fiore}, L. and {Di Lieto}, A. and {Emilio}, M. Di Paolo and {Di Virgilio}, A. and {Dietz}, A. and {Doets}, M. and {Dominici}, P. and {Dominjon}, A. and {Drago}, M. and {Drezen}, C. and {Dujardin}, B. and {Dulach}, B. and {Eder}, C. and {Eleuteri}, A. and {Enard}, D. and {Evans}, M. and {Fabbroni}, L. and {Fafone}, V. and {Fang}, H. and {Ferrante}, I. and {Fidecaro}, F. and {Fiori}, I. and {Flaminio}, R. and {Forest}, D. and {Forte}, L.~A. and {Fournier}, J. -D. and {Fournier}, L. and {Franc}, J. and {Francois}, O. and {Frasca}, S. and {Frasconi}, F. and {Freise}, A. and {Gaddi}, A. and {Galimberti}, M. and {Gammaitoni}, L. and {Ganau}, P. and {Garnier}, C. and {Garufi}, F. and {G{\'a}sp{\'a}r}, M.~E. and {Gemme}, G. and {Genin}, E. and {Gennai}, A. and {Gennaro}, G. and {Giacobone}, L. and {Giazotto}, A. and {Giordano}, G. and {Giordano}, L. and {Girard}, C. and {Gouaty}, R. and {Grado}, A. and {Granata}, M. and {Granata}, V. and {Grave}, X. and {Greverie}, C. and {Groenstege}, H. and {Guidi}, G.~M. and {Hamdani}, S. and {Hayau}, J. -F. and {Hebri}, S. and {Heidmann}, A. and {Heitmann}, H. and {Hello}, P. and {Hemming}, G. and {Hennes}, E. and {Hermel}, R. and {Heusse}, P. and {Holloway}, L. and {Huet}, D. and {Iannarelli}, M. and {Jaranowski}, P. and {Jehanno}, D. and {Journet}, L. and {Karkar}, S. and {Ketel}, T. and {Voet}, H. and {Kovalik}, J. and {Kowalska}, I. and {Kreckelbergh}, S. and {Krolak}, A. and {Lacotte}, J.~C. and {Lagrange}, B. and {La Penna}, P. and {Laval}, M. and {Le Marec}, J.~C. and {Leroy}, N. and {Letendre}, N. and {Li}, T.~G.~F. and {Lieunard}, B. and {Liguori}, N. and {Lodygensky}, O. and {Lopez}, B. and {Lorenzini}, M. and {Loriette}, V. and {Losurdo}, G. and {Loupias}, M. and {Mackowski}, J.~M.},
        title = "{Virgo: a laser interferometer to detect gravitational waves}",
      journal = {Journal of Instrumentation},
         year = 2012,
        month = mar,
       volume = {7},
       number = {3},
        pages = {3012},
          doi = {10.1088/1748-0221/7/03/P03012},
       adsurl = {https://ui.adsabs.harvard.edu/abs/2012JInst...7.3012A}
}

@ARTICLE{2015CQGra..32b4001A,
       author = {{Acernese}, F. and {Agathos}, M. and {Agatsuma}, K. and {Aisa}, D. and {Allemandou}, N. and {Allocca}, A. and {Amarni}, J. and {Astone}, P. and {Balestri}, G. and {Ballardin}, G. and {Barone}, F. and {Baronick}, J. -P. and {Barsuglia}, M. and {Basti}, A. and {Basti}, F. and {Bauer}, Th S. and {Bavigadda}, V. and {Bejger}, M. and {Beker}, M.~G. and {Belczynski}, C. and {Bersanetti}, D. and {Bertolini}, A. and {Bitossi}, M. and {Bizouard}, M.~A. and {Bloemen}, S. and {Blom}, M. and {Boer}, M. and {Bogaert}, G. and {Bondi}, D. and {Bondu}, F. and {Bonelli}, L. and {Bonnand}, R. and {Boschi}, V. and {Bosi}, L. and {Bouedo}, T. and {Bradaschia}, C. and {Branchesi}, M. and {Briant}, T. and {Brillet}, A. and {Brisson}, V. and {Bulik}, T. and {Bulten}, H.~J. and {Buskulic}, D. and {Buy}, C. and {Cagnoli}, G. and {Calloni}, E. and {Campeggi}, C. and {Canuel}, B. and {Carbognani}, F. and {Cavalier}, F. and {Cavalieri}, R. and {Cella}, G. and {Cesarini}, E. and {Mottin}, E. Chassande- and {Chincarini}, A. and {Chiummo}, A. and {Chua}, S. and {Cleva}, F. and {Coccia}, E. and {Cohadon}, P. -F. and {Colla}, A. and {Colombini}, M. and {Conte}, A. and {Coulon}, J. -P. and {Cuoco}, E. and {Dalmaz}, A. and {D'Antonio}, S. and {Dattilo}, V. and {Davier}, M. and {Day}, R. and {Debreczeni}, G. and {Degallaix}, J. and {Del{\'e}glise}, S. and {Pozzo}, W. Del and {Dereli}, H. and {Rosa}, R. De and {Fiore}, L. Di and {Lieto}, A. Di and {Virgilio}, A. Di and {Doets}, M. and {Dolique}, V. and {Drago}, M. and {Ducrot}, M. and {Endr{\H{o}}czi}, G. and {Fafone}, V. and {Farinon}, S. and {Ferrante}, I. and {Ferrini}, F. and {Fidecaro}, F. and {Fiori}, I. and {Flaminio}, R. and {Fournier}, J. -D. and {Franco}, S. and {Frasca}, S. and {Frasconi}, F. and {Gammaitoni}, L. and {Garufi}, F. and {Gaspard}, M. and {Gatto}, A. and {Gemme}, G. and {Gendre}, B. and {Genin}, E. and {Gennai}, A. and {Ghosh}, S. and {Giacobone}, L. and {Giazotto}, A. and {Gouaty}, R. and {Granata}, M. and {Greco}, G. and {Groot}, P. and {Guidi}, G.~M. and {Harms}, J. and {Heidmann}, A. and {Heitmann}, H. and {Hello}, P. and {Hemming}, G. and {Hennes}, E. and {Hofman}, D. and {Jaranowski}, P. and {Jonker}, R.~J.~G. and {Kasprzack}, M. and {K{\'e}f{\'e}lian}, F. and {Kowalska}, I. and {Kraan}, M. and {Kr{\'o}lak}, A. and {Kutynia}, A. and {Lazzaro}, C. and {Leonardi}, M. and {Leroy}, N. and {Letendre}, N. and {Li}, T.~G.~F. and {Lieunard}, B. and {Lorenzini}, M. and {Loriette}, V. and {Losurdo}, G. and {Magazz{\`u}}, C. and {Majorana}, E. and {Maksimovic}, I. and {Malvezzi}, V. and {Man}, N. and {Mangano}, V. and {Mantovani}, M. and {Marchesoni}, F. and {Marion}, F. and {Marque}, J. and {Martelli}, F. and {Martellini}, L. and {Masserot}, A. and {Meacher}, D. and {Meidam}, J. and {Mezzani}, F. and {Michel}, C. and {Milano}, L. and {Minenkov}, Y. and {Moggi}, A. and {Mohan}, M. and {Montani}, M. and {Morgado}, N. and {Mours}, B. and {Mul}, F. and {Nagy}, M.~F. and {Nardecchia}, I. and {Naticchioni}, L. and {Nelemans}, G. and {Neri}, I. and {Neri}, M. and {Nocera}, F. and {Pacaud}, E. and {Palomba}, C. and {Paoletti}, F. and {Paoli}, A. and {Pasqualetti}, A. and {Passaquieti}, R. and {Passuello}, D. and {Perciballi}, M. and {Petit}, S. and {Pichot}, M. and {Piergiovanni}, F. and {Pillant}, G. and {Piluso}, A. and {Pinard}, L. and {Poggiani}, R. and {Prijatelj}, M. and {Prodi}, G.~A. and {Punturo}, M. and {Puppo}, P. and {Rabeling}, D.~S. and {R{\'a}cz}, I. and {Rapagnani}, P. and {Razzano}, M. and {Re}, V. and {Regimbau}, T. and {Ricci}, F. and {Robinet}, F. and {Rocchi}, A. and {Rolland}, L. and {Romano}, R. and {Rosi{\'n}ska}, D. and {Ruggi}, P. and {Saracco}, E.},
        title = "{Advanced Virgo: a second-generation interferometric gravitational wave detector}",
      journal = {Classical and Quantum Gravity},
         year = 2015,
        month = jan,
       volume = {32},
       number = {2},
          eid = {024001},
        pages = {024001},
          doi = {10.1088/0264-9381/32/2/024001},
archivePrefix = {arXiv},
       eprint = {1408.3978},
 primaryClass = {gr-qc},
       adsurl = {https://ui.adsabs.harvard.edu/abs/2015CQGra..32b4001A}
}

@ARTICLE{2016PhRvD..93k2004M,
       author = {{Martynov}, D.~V. and {Hall}, E.~D. and {Abbott}, B.~P. and {Abbott}, R. and {Abbott}, T.~D. and {Adams}, C. and {Adhikari}, R.~X. and {Anderson}, R.~A. and {Anderson}, S.~B. and {Arai}, K. and {Arain}, M.~A. and {Aston}, S.~M. and {Austin}, L. and {Ballmer}, S.~W. and {Barbet}, M. and {Barker}, D. and {Barr}, B. and {Barsotti}, L. and {Bartlett}, J. and {Barton}, M.~A. and {Bartos}, I. and {Batch}, J.~C. and {Bell}, A.~S. and {Belopolski}, I. and {Bergman}, J. and {Betzwieser}, J. and {Billingsley}, G. and {Birch}, J. and {Biscans}, S. and {Biwer}, C. and {Black}, E. and {Blair}, C.~D. and {Bogan}, C. and {Bork}, R. and {Bridges}, D.~O. and {Brooks}, A.~F. and {Celerier}, C. and {Ciani}, G. and {Clara}, F. and {Cook}, D. and {Countryman}, S.~T. and {Cowart}, M.~J. and {Coyne}, D.~C. and {Cumming}, A. and {Cunningham}, L. and {Damjanic}, M. and {Dannenberg}, R. and {Danzmann}, K. and {Costa}, C.~F. Da Silva and {Daw}, E.~J. and {DeBra}, D. and {DeRosa}, R.~T. and {DeSalvo}, R. and {Dooley}, K.~L. and {Doravari}, S. and {Driggers}, J.~C. and {Dwyer}, S.~E. and {Effler}, A. and {Etzel}, T. and {Evans}, M. and {Evans}, T.~M. and {Factourovich}, M. and {Fair}, H. and {Feldbaum}, D. and {Fisher}, R.~P. and {Foley}, S. and {Frede}, M. and {Fritschel}, P. and {Frolov}, V.~V. and {Fulda}, P. and {Fyffe}, M. and {Galdi}, V. and {Giaime}, J.~A. and {Giardina}, K.~D. and {Gleason}, J.~R. and {Goetz}, R. and {Gras}, S. and {Gray}, C. and {Greenhalgh}, R.~J.~S. and {Grote}, H. and {Guido}, C.~J. and {Gushwa}, K.~E. and {Gustafson}, E.~K. and {Gustafson}, R. and {Hammond}, G. and {Hanks}, J. and {Hanson}, J. and {Hardwick}, T. and {Harry}, G.~M. and {Heefner}, J. and {Heintze}, M.~C. and {Heptonstall}, A.~W. and {Hoak}, D. and {Hough}, J. and {Ivanov}, A. and {Izumi}, K. and {Jacobson}, M. and {James}, E. and {Jones}, R. and {Kandhasamy}, S. and {Karki}, S. and {Kasprzack}, M. and {Kaufer}, S. and {Kawabe}, K. and {Kells}, W. and {Kijbunchoo}, N. and {King}, E.~J. and {King}, P.~J. and {Kinzel}, D.~L. and {Kissel}, J.~S. and {Kokeyama}, K. and {Korth}, W.~Z. and {Kuehn}, G. and {Kwee}, P. and {Landry}, M. and {Lantz}, B. and {Le Roux}, A. and {Levine}, B.~M. and {Lewis}, J.~B. and {Lhuillier}, V. and {Lockerbie}, N.~A. and {Lormand}, M. and {Lubinski}, M.~J. and {Lundgren}, A.~P. and {MacDonald}, T. and {MacInnis}, M. and {Macleod}, D.~M. and {Mageswaran}, M. and {Mailand}, K. and {M{\'a}rka}, S. and {M{\'a}rka}, Z. and {Markosyan}, A.~S. and {Maros}, E. and {Martin}, I.~W. and {Martin}, R.~M. and {Marx}, J.~N. and {Mason}, K. and {Massinger}, T.~J. and {Matichard}, F. and {Mavalvala}, N. and {McCarthy}, R. and {McClelland}, D.~E. and {McCormick}, S. and {McIntyre}, G. and {McIver}, J. and {Merilh}, E.~L. and {Meyer}, M.~S. and {Meyers}, P.~M. and {Miller}, J. and {Mittleman}, R. and {Moreno}, G. and {Mueller}, C.~L. and {Mueller}, G. and {Mullavey}, A. and {Munch}, J. and {Nuttall}, L.~K. and {Oberling}, J. and {O'Dell}, J. and {Oppermann}, P. and {Oram}, Richard J. and {O'Reilly}, B. and {Osthelder}, C. and {Ottaway}, D.~J. and {Overmier}, H. and {Palamos}, J.~R. and {Paris}, H.~R. and {Parker}, W. and {Patrick}, Z. and {Pele}, A. and {Penn}, S. and {Phelps}, M. and {Pickenpack}, M. and {Pierro}, V. and {Pinto}, I. and {Poeld}, J. and {Principe}, M. and {Prokhorov}, L. and {Puncken}, O. and {Quetschke}, V. and {Quintero}, E.~A. and {Raab}, F.~J. and {Radkins}, H. and {Raffai}, P. and {Ramet}, C.~R. and {Reed}, C.~M. and {Reid}, S. and {Reitze}, D.~H. and {Robertson}, N.~A. and {Rollins}, J.~G. and {Roma}, V.~J. and {Romie}, J.~H. and {Rowan}, S. and {Ryan}, K. and {Sadecki}, T. and {Sanchez}, E.~J. and {Sandberg}, V. and {Sannibale}, V. and {Savage}, R.~L. and {Schofield}, R.~M.~S. and {Schultz}, B.},
        title = "{Sensitivity of the Advanced LIGO detectors at the beginning of gravitational wave astronomy}",
      journal = {\prd},
         year = 2016,
        month = jun,
       volume = {93},
       number = {11},
          eid = {112004},
        pages = {112004},
          doi = {10.1103/PhysRevD.93.112004},
archivePrefix = {arXiv},
       eprint = {1604.00439},
 primaryClass = {astro-ph.IM},
       adsurl = {https://ui.adsabs.harvard.edu/abs/2016PhRvD..93k2004M}
}

@ARTICLE{2019NatAs...3...35K,
       author = {{Kagra Collaboration} and {Akutsu}, T. and {Ando}, M. and {Arai}, K. and {Arai}, Y. and {Araki}, S. and {Araya}, A. and {Aritomi}, N. and {Asada}, H. and {Aso}, Y. and {Atsuta}, S. and {Awai}, K. and {Bae}, S. and {Baiotti}, L. and {Barton}, M.~A. and {Cannon}, K. and {Capocasa}, E. and {Chen}, C. -S. and {Chiu}, T. -W. and {Cho}, K. and {Chu}, Y. -K. and {Craig}, K. and {Creus}, W. and {Doi}, K. and {Eda}, K. and {Enomoto}, Y. and {Flaminio}, R. and {Fujii}, Y. and {Fujimoto}, M. -K. and {Fukunaga}, M. and {Fukushima}, M. and {Furuhata}, T. and {Haino}, S. and {Hasegawa}, K. and {Hashino}, K. and {Hayama}, K. and {Hirobayashi}, S. and {Hirose}, E. and {Hsieh}, B.~H. and {Huang}, C. -Z. and {Ikenoue}, B. and {Inoue}, Y. and {Ioka}, K. and {Itoh}, Y. and {Izumi}, K. and {Kaji}, T. and {Kajita}, T. and {Kakizaki}, M. and {Kamiizumi}, M. and {Kanbara}, S. and {Kanda}, N. and {Kanemura}, S. and {Kaneyama}, M. and {Kang}, G. and {Kasuya}, J. and {Kataoka}, Y. and {Kawai}, N. and {Kawamura}, S. and {Kawasaki}, T. and {Kim}, C. and {Kim}, J. and {Kim}, J.~C. and {Kim}, W.~S. and {Kim}, Y. -M. and {Kimura}, N. and {Kinugawa}, T. and {Kirii}, S. and {Kitaoka}, Y. and {Kitazawa}, H. and {Kojima}, Y. and {Kokeyama}, K. and {Komori}, K. and {Kong}, A.~K.~H. and {Kotake}, K. and {Kozu}, R. and {Kumar}, R. and {Kuo}, H. -S. and {Kuroyanagi}, S. and {Lee}, H.~K. and {Lee}, H.~M. and {Lee}, H.~W. and {Leonardi}, M. and {Lin}, C. -Y. and {Lin}, F. -L. and {Liu}, G.~C. and {Liu}, Y. and {Majorana}, E. and {Mano}, S. and {Marchio}, M. and {Matsui}, T. and {Matsushima}, F. and {Michimura}, Y. and {Mio}, N. and {Miyakawa}, O. and {Miyamoto}, A. and {Miyamoto}, T. and {Miyo}, K. and {Miyoki}, S. and {Morii}, W. and {Morisaki}, S. and {Moriwaki}, Y. and {Morozumi}, T. and {Musha}, M. and {Nagano}, K. and {Nagano}, S. and {Nakamura}, K. and {Nakamura}, T. and {Nakano}, H. and {Nakano}, M. and {Nakao}, K. and {Narikawa}, T. and {Naticchioni}, L. and {Nguyen Quynh}, L. and {Ni}, W. -T. and {Nishizawa}, A. and {Obuchi}, Y. and {Ochi}, T. and {Oh}, J.~J. and {Oh}, S.~H. and {Ohashi}, M. and {Ohishi}, N. and {Ohkawa}, M. and {Okutomi}, K. and {Ono}, K. and {Oohara}, K. and {Ooi}, C.~P. and {Pan}, S. -S. and {Park}, J. and {Pe{\~n}a Arellano}, F.~E. and {Pinto}, I. and {Sago}, N. and {Saijo}, M. and {Saitou}, S. and {Saito}, Y. and {Sakai}, K. and {Sakai}, Y. and {Sakai}, Y. and {Sasai}, M. and {Sasaki}, M. and {Sasaki}, Y. and {Sato}, S. and {Sato}, N. and {Sato}, T. and {Sekiguchi}, Y. and {Seto}, N. and {Shibata}, M. and {Shimoda}, T. and {Shinkai}, H. and {Shishido}, T. and {Shoda}, A. and {Somiya}, K. and {Son}, E.~J. and {Suemasa}, A. and {Suzuki}, T. and {Suzuki}, T. and {Tagoshi}, H. and {Tahara}, H. and {Takahashi}, H. and {Takahashi}, R. and {Takamori}, A. and {Takeda}, H. and {Tanaka}, H. and {Tanaka}, K. and {Tanaka}, T. and {Tanioka}, S. and {Tapia San Martin}, E.~N. and {Tatsumi}, D. and {Tomaru}, T. and {Tomura}, T. and {Travasso}, F. and {Tsubono}, K. and {Tsuchida}, S. and {Uchikata}, N. and {Uchiyama}, T. and {Uehara}, T. and {Ueki}, S. and {Ueno}, K. and {Uraguchi}, F. and {Ushiba}, T. and {van Putten}, M.~H.~P.~M. and {Vocca}, H. and {Wada}, S. and {Wakamatsu}, T. and {Watanabe}, Y. and {Xu}, W. -R. and {Yamada}, T. and {Yamamoto}, A. and {Yamamoto}, K. and {Yamamoto}, K. and {Yamamoto}, S. and {Yamamoto}, T. and {Yokogawa}, K. and {Yokoyama}, J. and {Yokozawa}, T. and {Yoon}, T.~H. and {Yoshioka}, T. and {Yuzurihara}, H. and {Zeidler}, S. and {Zhu}, Z. -H.},
        title = "{KAGRA: 2.5 generation interferometric gravitational wave detector}",
      journal = {Nature Astronomy},
         year = 2019,
        month = jan,
       volume = {3},
        pages = {35-40},
          doi = {10.1038/s41550-018-0658-y},
archivePrefix = {arXiv},
       eprint = {1811.08079},
 primaryClass = {gr-qc},
       adsurl = {https://ui.adsabs.harvard.edu/abs/2019NatAs...3...35K}
}

@ARTICLE{2021PTEP.2021eA101A,
       author = {{Akutsu}, T. and {Ando}, M. and {Arai}, K. and {Arai}, Y. and {Araki}, S. and {Araya}, A. and {Aritomi}, N. and {Aso}, Y. and {Bae}, S. and {Bae}, Y. and {Baiotti}, L. and {Bajpai}, R. and {Barton}, M.~A. and {Cannon}, K. and {Capocasa}, E. and {Chan}, M. and {Chen}, C. and {Chen}, K. and {Chen}, Y. and {Chu}, H. and {Chu}, Y. -K. and {Eguchi}, S. and {Enomoto}, Y. and {Flaminio}, R. and {Fujii}, Y. and {Fukunaga}, M. and {Fukushima}, M. and {Ge}, G. and {Hagiwara}, A. and {Haino}, S. and {Hasegawa}, K. and {Hayakawa}, H. and {Hayama}, K. and {Himemoto}, Y. and {Hiranuma}, Y. and {Hirata}, N. and {Hirose}, E. and {Hong}, Z. and {Hsieh}, B.~H. and {Huang}, C. -Z. and {Huang}, P. and {Huang}, Y. and {Ikenoue}, B. and {Imam}, S. and {Inayoshi}, K. and {Inoue}, Y. and {Ioka}, K. and {Itoh}, Y. and {Izumi}, K. and {Jung}, K. and {Jung}, P. and {Kajita}, T. and {Kamiizumi}, M. and {Kanda}, N. and {Kang}, G. and {Kawaguchi}, K. and {Kawai}, N. and {Kawasaki}, T. and {Kim}, C. and {Kim}, J.~C. and {Kim}, W.~S. and {Kim}, Y. -M. and {Kimura}, N. and {Kita}, N. and {Kitazawa}, H. and {Kojima}, Y. and {Kokeyama}, K. and {Komori}, K. and {Kong}, A.~K.~H. and {Kotake}, K. and {Kozakai}, C. and {Kozu}, R. and {Kumar}, R. and {Kume}, J. and {Kuo}, C. and {Kuo}, H. -S. and {Kuroyanagi}, S. and {Kusayanagi}, K. and {Kwak}, K. and {Lee}, H.~K. and {Lee}, H.~W. and {Lee}, R. and {Leonardi}, M. and {Lin}, L.~C. -C. and {Lin}, C. -Y. and {Lin}, F. -L. and {Liu}, G.~C. and {Luo}, L. -W. and {Marchio}, M. and {Michimura}, Y. and {Mio}, N. and {Miyakawa}, O. and {Miyamoto}, A. and {Miyazaki}, Y. and {Miyo}, K. and {Miyoki}, S. and {Morisaki}, S. and {Moriwaki}, Y. and {Nagano}, K. and {Nagano}, S. and {Nakamura}, K. and {Nakano}, H. and {Nakano}, M. and {Nakashima}, R. and {Narikawa}, T. and {Negishi}, R. and {Ni}, W. -T. and {Nishizawa}, A. and {Obuchi}, Y. and {Ogaki}, W. and {Oh}, J.~J. and {Oh}, S.~H. and {Ohashi}, M. and {Ohishi}, N. and {Ohkawa}, M. and {Okutomi}, K. and {Oohara}, K. and {Ooi}, C.~P. and {Oshino}, S. and {Pan}, K. and {Pang}, H. and {Park}, J. and {Arellano}, F.~E. Pe{\~n}a and {Pinto}, I. and {Sago}, N. and {Saito}, S. and {Saito}, Y. and {Sakai}, K. and {Sakai}, Y. and {Sakuno}, Y. and {Sato}, S. and {Sato}, T. and {Sawada}, T. and {Sekiguchi}, T. and {Sekiguchi}, Y. and {Shibagaki}, S. and {Shimizu}, R. and {Shimoda}, T. and {Shimode}, K. and {Shinkai}, H. and {Shishido}, T. and {Shoda}, A. and {Somiya}, K. and {Son}, E.~J. and {Sotani}, H. and {Sugimoto}, R. and {Suzuki}, T. and {Suzuki}, T. and {Tagoshi}, H. and {Takahashi}, H. and {Takahashi}, R. and {Takamori}, A. and {Takano}, S. and {Takeda}, H. and {Takeda}, M. and {Tanaka}, H. and {Tanaka}, K. and {Tanaka}, K. and {Tanaka}, T. and {Tanaka}, T. and {Tanioka}, S. and {Tapia San Martin}, E.~N. and {Telada}, S. and {Tomaru}, T. and {Tomigami}, Y. and {Tomura}, T. and {Travasso}, F. and {Trozzo}, L. and {Tsang}, T. and {Tsubono}, K. and {Tsuchida}, S. and {Tsuzuki}, T. and {Tuyenbayev}, D. and {Uchikata}, N. and {Uchiyama}, T. and {Ueda}, A. and {Uehara}, T. and {Ueno}, K. and {Ueshima}, G. and {Uraguchi}, F. and {Ushiba}, T. and {van Putten}, M.~H.~P.~M. and {Vocca}, H. and {Wang}, J. and {Wu}, C. and {Wu}, H. and {Wu}, S. and {Xu}, W. -R. and {Yamada}, T. and {Yamamoto}, K. and {Yamamoto}, K. and {Yamamoto}, T. and {Yokogawa}, K. and {Yokoyama}, J. and {Yokozawa}, T. and {Yoshioka}, T. and {Yuzurihara}, H. and {Zeidler}, S. and {Zhao}, Y. and {Zhu}, Z. -H.},
        title = "{Overview of KAGRA: Detector design and construction history}",
      journal = {Progress of Theoretical and Experimental Physics},
         year = 2021,
        month = may,
       volume = {2021},
       number = {5},
          eid = {05A101},
        pages = {05A101},
          doi = {10.1093/ptep/ptaa125},
archivePrefix = {arXiv},
       eprint = {2005.05574},
 primaryClass = {physics.ins-det},
       adsurl = {https://ui.adsabs.harvard.edu/abs/2021PTEP.2021eA101A}
}

@ARTICLE{2012JAI.....150004T,
       author = {{Taylor}, G.~B. and {Ellingson}, S.~W. and {Kassim}, N.~E. and {Craig}, J. and {Dowell}, J. and {Wolfe}, C.~N. and {Hartman}, J. and {Bernardi}, G. and {Clarke}, T. and {Cohen}, A. and {Dalal}, N.~P. and {Erickson}, W.~C. and {Hicks}, B. and {Greenhill}, L.~J. and {Jacoby}, B. and {Lane}, W. and {Lazio}, J. and {Mitchell}, D. and {Navarro}, R. and {Ord}, S.~M. and {Pihlstr{\"o}m}, Y. and {Polisensky}, E. and {Ray}, P.~S. and {Rickard}, L.~J. and {Schinzel}, F.~K. and {Schmitt}, H. and {Sigman}, E. and {Soriano}, M. and {Stewart}, K.~P. and {Stovall}, K. and {Tremblay}, S. and {Wang}, D. and {Weiler}, K.~W. and {White}, S. and {Wood}, D.~L.},
        title = "{First Light for the First Station of the Long Wavelength Array}",
      journal = {Journal of Astronomical Instrumentation},
         year = 2012,
        month = dec,
       volume = {1},
       number = {1},
          eid = {1250004-284},
        pages = {1250004-284},
          doi = {10.1142/S2251171712500043},
archivePrefix = {arXiv},
       eprint = {1206.6733},
 primaryClass = {astro-ph.IM},
       adsurl = {https://ui.adsabs.harvard.edu/abs/2012JAI.....150004T}
}

@ARTICLE{2019ApJ...886..123A,
       author = {{Anderson}, Marin M. and {Hallinan}, Gregg and {Eastwood}, Michael W. and {Monroe}, Ryan M. and {Callister}, Thomas A. and {Dowell}, Jayce and {Hicks}, Brian and {Huang}, Yuping and {Kassim}, Namir E. and {Kocz}, Jonathon and {Lazio}, T. Joseph W. and {Price}, Danny C. and {Schinzel}, Frank K. and {Taylor}, Greg B.},
        title = "{New Limits on the Low-frequency Radio Transient Sky Using 31 hr of All-sky Data with the OVRO-LWA}",
      journal = {\apj},
         year = 2019,
        month = dec,
       volume = {886},
       number = {2},
          eid = {123},
        pages = {123},
          doi = {10.3847/1538-4357/ab4f87},
archivePrefix = {arXiv},
       eprint = {1911.04591},
 primaryClass = {astro-ph.HE},
       adsurl = {https://ui.adsabs.harvard.edu/abs/2019ApJ...886..123A}
}

@ARTICLE{2020NIMPA.95363086M,
       author = {{Monroe}, Ryan and {Romero Wolf}, Andres and {Hallinan}, Gregg and {Nelles}, Anna and {Eastwood}, Michael and {Anderson}, Marin and {D'Addario}, Larry and {Kocz}, Jonathon and {Wang}, Yuankun and {Cody}, Devin and {Woody}, David and {Schinzel}, Frank and {Taylor}, Greg and {Greenhill}, Lincoln and {Price}, Daniel},
        title = "{Self-triggered radio detection and identification of cosmic air showers with the OVRO-LWA}",
      journal = {Nuclear Instruments and Methods in Physics Research A},
         year = 2020,
        month = feb,
       volume = {953},
          eid = {163086},
        pages = {163086},
          doi = {10.1016/j.nima.2019.163086},
archivePrefix = {arXiv},
       eprint = {1907.10193},
 primaryClass = {astro-ph.IM},
       adsurl = {https://ui.adsabs.harvard.edu/abs/2020NIMPA.95363086M}
}

@INPROCEEDINGS{2022icrc.confE.204P,
       author = {{Plant}, K. and {Romero-Wolf}, A. and {Carvalho}, W. and {Belov}, K. and {Hallinan}, G.},
        title = "{Updates from the OVRO-LWA: Commissioning a Full-Duty-Cycle Radio-Only Cosmic Ray Detector}",
    booktitle = {37th International Cosmic Ray Conference},
         year = 2022,
        month = mar,
          eid = {204},
        pages = {204},
          doi = {10.22323/1.395.0204},
       adsurl = {https://ui.adsabs.harvard.edu/abs/2022icrc.confE.204P}
}

@ARTICLE{2018MNRAS.478.4193P,
       author = {{Price}, D.~C. and {Greenhill}, L.~J. and {Fialkov}, A. and {Bernardi}, G. and {Garsden}, H. and {Barsdell}, B.~R. and {Kocz}, J. and {Anderson}, M.~M. and {Bourke}, S.~A. and {Craig}, J. and {Dexter}, M.~R. and {Dowell}, J. and {Eastwood}, M.~W. and {Eftekhari}, T. and {Ellingson}, S.~W. and {Hallinan}, G. and {Hartman}, J.~M. and {Kimberk}, R. and {Lazio}, T. Joseph W. and {Leiker}, S. and {MacMahon}, D. and {Monroe}, R. and {Schinzel}, F. and {Taylor}, G.~B. and {Tong}, E. and {Werthimer}, D. and {Woody}, D.~P.},
        title = "{Design and characterization of the Large-aperture Experiment to Detect the Dark Age (LEDA) radiometer systems}",
      journal = {\mnras},
         year = 2018,
        month = aug,
       volume = {478},
       number = {3},
        pages = {4193-4213},
          doi = {10.1093/mnras/sty1244},
archivePrefix = {arXiv},
       eprint = {1709.09313},
 primaryClass = {astro-ph.IM},
       adsurl = {https://ui.adsabs.harvard.edu/abs/2018MNRAS.478.4193P}
}

@ARTICLE{2019AJ....158...84E,
       author = {{Eastwood}, Michael W. and {Anderson}, Marin M. and {Monroe}, Ryan M. and {Hallinan}, Gregg and {Catha}, Morgan and {Dowell}, Jayce and {Garsden}, Hugh and {Greenhill}, Lincoln J. and {Hicks}, Brian C. and {Kocz}, Jonathon and {Price}, Danny C. and {Schinzel}, Frank K. and {Vedantham}, Harish and {Wang}, Yuankun},
        title = "{The 21 cm Power Spectrum from the Cosmic Dawn: First Results from the OVRO-LWA}",
      journal = {\aj},
         year = 2019,
        month = aug,
       volume = {158},
       number = {2},
          eid = {84},
        pages = {84},
          doi = {10.3847/1538-3881/ab2629},
archivePrefix = {arXiv},
       eprint = {1906.08943},
 primaryClass = {astro-ph.CO},
       adsurl = {https://ui.adsabs.harvard.edu/abs/2019AJ....158...84E}
}

@ARTICLE{2021MNRAS.506.5802G,
       author = {{Garsden}, H. and {Greenhill}, L. and {Bernardi}, G. and {Fialkov}, A. and {Price}, D.~C. and {Mitchell}, D. and {Dowell}, J. and {Spinelli}, M. and {Schinzel}, F.~K.},
        title = "{A 21-cm power spectrum at 48 MHz, using the Owens Valley Long Wavelength Array}",
      journal = {\mnras},
         year = 2021,
        month = oct,
       volume = {506},
       number = {4},
        pages = {5802-5817},
          doi = {10.1093/mnras/stab1671},
archivePrefix = {arXiv},
       eprint = {2102.09596},
 primaryClass = {astro-ph.CO},
       adsurl = {https://ui.adsabs.harvard.edu/abs/2021MNRAS.506.5802G}
}

@ARTICLE{2014MNRAS.444..606O,
       author = {{Offringa}, A.~R. and {McKinley}, B. and {Hurley-Walker}, N. and {Briggs}, F.~H. and {Wayth}, R.~B. and {Kaplan}, D.~L. and {Bell}, M.~E. and {Feng}, L. and {Neben}, A.~R. and {Hughes}, J.~D. and {Rhee}, J. and {Murphy}, T. and {Bhat}, N.~D.~R. and {Bernardi}, G. and {Bowman}, J.~D. and {Cappallo}, R.~J. and {Corey}, B.~E. and {Deshpande}, A.~A. and {Emrich}, D. and {Ewall-Wice}, A. and {Gaensler}, B.~M. and {Goeke}, R. and {Greenhill}, L.~J. and {Hazelton}, B.~J. and {Hindson}, L. and {Johnston-Hollitt}, M. and {Jacobs}, D.~C. and {Kasper}, J.~C. and {Kratzenberg}, E. and {Lenc}, E. and {Lonsdale}, C.~J. and {Lynch}, M.~J. and {McWhirter}, S.~R. and {Mitchell}, D.~A. and {Morales}, M.~F. and {Morgan}, E. and {Kudryavtseva}, N. and {Oberoi}, D. and {Ord}, S.~M. and {Pindor}, B. and {Procopio}, P. and {Prabu}, T. and {Riding}, J. and {Roshi}, D.~A. and {Shankar}, N. Udaya and {Srivani}, K.~S. and {Subrahmanyan}, R. and {Tingay}, S.~J. and {Waterson}, M. and {Webster}, R.~L. and {Whitney}, A.~R. and {Williams}, A. and {Williams}, C.~L.},
        title = "{WSCLEAN: an implementation of a fast, generic wide-field imager for radio astronomy}",
      journal = {\mnras},
         year = 2014,
        month = oct,
       volume = {444},
       number = {1},
        pages = {606-619},
          doi = {10.1093/mnras/stu1368},
archivePrefix = {arXiv},
       eprint = {1407.1943},
 primaryClass = {astro-ph.IM},
       adsurl = {https://ui.adsabs.harvard.edu/abs/2014MNRAS.444..606O}
}

@software{2016zndo...1049160E,
       author = {{Eastwood}, Michael W.},
        title = "{TTCal}",
         year = 2016,
        month = oct,
          eid = {10.5281/zenodo.1049160},
          doi = {10.5281/zenodo.1049160},
      version = {0.3.0},
    publisher = {Zenodo},
       adsurl = {https://ui.adsabs.harvard.edu/abs/2016zndo...1049160E}
}

@ARTICLE{2022PASP..134k4501C,
       author = {{CASA Team} and {Bean}, Ben and {Bhatnagar}, Sanjay and {Castro}, Sandra and {Donovan Meyer}, Jennifer and {Emonts}, Bjorn and {Garcia}, Enrique and {Garwood}, Robert and {Golap}, Kumar and {Gonzalez Villalba}, Justo and {Harris}, Pamela and {Hayashi}, Yohei and {Hoskins}, Josh and {Hsieh}, Mingyu and {Jagannathan}, Preshanth and {Kawasaki}, Wataru and {Keimpema}, Aard and {Kettenis}, Mark and {Lopez}, Jorge and {Marvil}, Joshua and {Masters}, Joseph and {McNichols}, Andrew and {Mehringer}, David and {Miel}, Renaud and {Moellenbrock}, George and {Montesino}, Federico and {Nakazato}, Takeshi and {Ott}, Juergen and {Petry}, Dirk and {Pokorny}, Martin and {Raba}, Ryan and {Rau}, Urvashi and {Schiebel}, Darrell and {Schweighart}, Neal and {Sekhar}, Srikrishna and {Shimada}, Kazuhiko and {Small}, Des and {Steeb}, Jan-Willem and {Sugimoto}, Kanako and {Suoranta}, Ville and {Tsutsumi}, Takahiro and {van Bemmel}, Ilse M. and {Verkouter}, Marjolein and {Wells}, Akeem and {Xiong}, Wei and {Szomoru}, Arpad and {Griffith}, Morgan and {Glendenning}, Brian and {Kern}, Jeff},
        title = "{CASA, the Common Astronomy Software Applications for Radio Astronomy}",
      journal = {\pasp},
         year = 2022,
        month = nov,
       volume = {134},
       number = {1041},
          eid = {114501},
        pages = {114501},
          doi = {10.1088/1538-3873/ac9642},
archivePrefix = {arXiv},
       eprint = {2210.02276},
 primaryClass = {astro-ph.IM},
       adsurl = {https://ui.adsabs.harvard.edu/abs/2022PASP..134k4501C}
}

@software{2010ascl.soft10017O,
       author = {{Offringa}, A.~R.},
        title = "{AOFlagger: RFI Software}",
 howpublished = {Astrophysics Source Code Library, record ascl:1010.017},
         year = 2010,
        month = oct,
          eid = {ascl:1010.017},
       adsurl = {https://ui.adsabs.harvard.edu/abs/2010ascl.soft10017O}
}

@ARTICLE{2025arXiv250700357A,
       author = {{Ahumada}, Tom{\'a}s and {Anand}, Shreya and {Bulla}, Mattia and {Gupta}, Vaidehi and {Kasliwal}, Mansi and {Stein}, Robert and {Karambelkar}, Viraj and {Bellm}, Eric C. and {Jegou du Laz}, Theophile and {Coughlin}, Michael W. and {Andreoni}, Igor and {Banerjee}, Smaranika and {Bochenek}, Aleksandra and {Hinds}, K-Ryan and {Hu}, Lei and {Palmese}, Antonella and {Perley}, Daniel and {Pletskova}, Natalya and {Salgundi}, Anirudh and {Singh}, Avinash and {Sollerman}, Jesper and {Swain}, Vishwajeet and {Wold}, Avery and {Bhalerao}, Varun and {Cenko}, S. Bradley and {Cook}, David O. and {Copperwheat}, Chris and {Graham}, Matthew and {Kaplan}, David L. and {Singer}, Leo P. and {Sravan}, Niharika and {Busmann}, Malte and {Gassert}, Julius and {Gruen}, Daniel and {Sommer}, Julian and {Zhang}, Yajie and {Amsellem}, Ariel and {Cabrera}, Tom{\'a}s and {Hall}, Xander J. and {Kunnumkai}, Keerthi and {O'Connor}, Brendan and {Barna}, Tyler and {Fontinele Nunes}, Felipe and {Toivonen}, Andrew and {Sasli}, Argyro and {Masci}, Frank J. and {Chen}, Tracy X. and {Dekany}, Richard and {Purdum}, Josiah and {Le-Calloch}, Antoine and {Anupama}, G.~C. and {Barway}, Sudhanshu},
        title = "{LIGO/Virgo/KAGRA neutron star merger candidate S250206dm: Zwicky Transient Facility observations}",
      journal = {arXiv e-prints},
         year = 2025,
        month = jul,
          eid = {arXiv:2507.00357},
        pages = {arXiv:2507.00357},
          doi = {10.48550/arXiv.2507.00357},
archivePrefix = {arXiv},
       eprint = {2507.00357},
 primaryClass = {astro-ph.HE},
       adsurl = {https://ui.adsabs.harvard.edu/abs/2025arXiv250700357A}
}

@ARTICLE{2025PASP..137g4203F,
       author = {{Frostig}, Danielle and {Karambelkar}, Viraj R. and {Stein}, Robert D. and {Lourie}, Nathan P. and {Kasliwal}, Mansi M. and {Simcoe}, Robert A. and {Bulla}, Mattia and {Ahumada}, Tom{\'a}s and {Mo}, Geoffrey and {Purdum}, Josiah and {Juneau}, Jill and {Malonis}, Andrew and {F{\H{u}}r{\'e}sz}, G{\'a}bor},
        title = "{WINTER on S250206dm: A Near-infrared Search for an Electromagnetic Counterpart to a Gravitational-wave Event}",
      journal = {\pasp},
         year = 2025,
        month = jul,
       volume = {137},
       number = {7},
          eid = {074203},
        pages = {074203},
          doi = {10.1088/1538-3873/ade478},
archivePrefix = {arXiv},
       eprint = {2504.12384},
 primaryClass = {astro-ph.HE},
       adsurl = {https://ui.adsabs.harvard.edu/abs/2025PASP..137g4203F}
}

@ARTICLE{2025GCN.39210....1Y,
       author = {{Young}, D.~R. and {Gillanders}, J.~H. and {Huber}, M.~E. and {Chambers}, K.~C. and {Smartt}, S.~J. and {Smith}, K.~W. and {Srivastav}, S. and {Stoppa}, F. and {Angus}, C.~R. and {Nicholl}, M. and {Fulton}, M.~D. and {McCollum}, M. and {Moore}, T. and {Sim}, S. and {Weston}, J. and {Aamer}, A. and {Sheng}, X. and {Ramsden}, P. and {Shingles}, L. and {Stevance}, H. and {''Oxford''} and {Rhodes}, L. and {Schultz}, A.~S.~B. and {de Boer}, T. and {Fairlamb}, J. and {Gao}, H. and {Lin}, C.~C. and {Lowe}, T. and {Magnier}, E. and {Minguez}, P. and {Paek}, G. and {Smith}, A. and {Wainscoat}, R.~J. and {Chen}, T. -W. and {Rest}, A. and {Stubbs}, C.},
        title = "{LIGO/Virgo/KAGRA S250206dm: Pan-STARRS survey of the skymap for optical transients}",
      journal = {GRB Coordinates Network},
         year = 2025,
        month = feb,
       volume = {39210},
        pages = {1},
       adsurl = {https://ui.adsabs.harvard.edu/abs/2025GCN.39210....1Y}
}

@ARTICLE{2025GCN.39176....1I,
       author = {{IceCube Collaboration}},
        title = "{LIGO/Virgo/KAGRA S250206dm: Two counterpart neutrino candidates from IceCube neutrino searches}",
      journal = {GRB Coordinates Network},
         year = 2025,
        month = feb,
       volume = {39176},
        pages = {1},
       adsurl = {https://ui.adsabs.harvard.edu/abs/2025GCN.39176....1I}
}

@ARTICLE{2025GCN.39216....1C,
       author = {{Chime/Frb Collaboration}},
        title = "{CHIME/FRB source FRB 20250206A detected less than 1 minute after LIGO-Virgo-Kagra (LVK) S250206dm, however the probability of spatial coincidence is order 0.1\%}",
      journal = {GRB Coordinates Network},
         year = 2025,
        month = feb,
       volume = {39216},
        pages = {1},
       adsurl = {https://ui.adsabs.harvard.edu/abs/2025GCN.39216....1C}
}

@ARTICLE{2019arXiv191201703P,
       author = {{Paszke}, Adam and {Gross}, Sam and {Massa}, Francisco and {Lerer}, Adam and {Bradbury}, James and {Chanan}, Gregory and {Killeen}, Trevor and {Lin}, Zeming and {Gimelshein}, Natalia and {Antiga}, Luca and {Desmaison}, Alban and {K{\"o}pf}, Andreas and {Yang}, Edward and {DeVito}, Zach and {Raison}, Martin and {Tejani}, Alykhan and {Chilamkurthy}, Sasank and {Steiner}, Benoit and {Fang}, Lu and {Bai}, Junjie and {Chintala}, Soumith},
        title = "{PyTorch: An Imperative Style, High-Performance Deep Learning Library}",
      journal = {arXiv e-prints},
         year = 2019,
        month = dec,
          eid = {arXiv:1912.01703},
        pages = {arXiv:1912.01703},
          doi = {10.48550/arXiv.1912.01703},
archivePrefix = {arXiv},
       eprint = {1912.01703},
 primaryClass = {cs.LG},
       adsurl = {https://ui.adsabs.harvard.edu/abs/2019arXiv191201703P}
}

@ARTICLE{2010PASP..122..595N,
       author = {{Nita}, Gelu M. and {Gary}, Dale E.},
        title = "{Statistics of the Spectral Kurtosis Estimator}",
      journal = {\pasp},
         year = 2010,
        month = may,
       volume = {122},
       number = {891},
        pages = {595},
          doi = {10.1086/652409},
       adsurl = {https://ui.adsabs.harvard.edu/abs/2010PASP..122..595N}
}

@ARTICLE{2017Sci...358.1565E,
       author = {{Evans}, P.~A. and {Cenko}, S.~B. and {Kennea}, J.~A. and {Emery}, S.~W.~K. and {Kuin}, N.~P.~M. and {Korobkin}, O. and {Wollaeger}, R.~T. and {Fryer}, C.~L. and {Madsen}, K.~K. and {Harrison}, F.~A. and {Xu}, Y. and {Nakar}, E. and {Hotokezaka}, K. and {Lien}, A. and {Campana}, S. and {Oates}, S.~R. and {Troja}, E. and {Breeveld}, A.~A. and {Marshall}, F.~E. and {Barthelmy}, S.~D. and {Beardmore}, A.~P. and {Burrows}, D.~N. and {Cusumano}, G. and {D'A{\`\i}}, A. and {D'Avanzo}, P. and {D'Elia}, V. and {de Pasquale}, M. and {Even}, W.~P. and {Fontes}, C.~J. and {Forster}, K. and {Garcia}, J. and {Giommi}, P. and {Grefenstette}, B. and {Gronwall}, C. and {Hartmann}, D.~H. and {Heida}, M. and {Hungerford}, A.~L. and {Kasliwal}, M.~M. and {Krimm}, H.~A. and {Levan}, A.~J. and {Malesani}, D. and {Melandri}, A. and {Miyasaka}, H. and {Nousek}, J.~A. and {O'Brien}, P.~T. and {Osborne}, J.~P. and {Pagani}, C. and {Page}, K.~L. and {Palmer}, D.~M. and {Perri}, M. and {Pike}, S. and {Racusin}, J.~L. and {Rosswog}, S. and {Siegel}, M.~H. and {Sakamoto}, T. and {Sbarufatti}, B. and {Tagliaferri}, G. and {Tanvir}, N.~R. and {Tohuvavohu}, A.},
        title = "{Swift and NuSTAR observations of GW170817: Detection of a blue kilonova}",
      journal = {Science},
         year = 2017,
        month = dec,
       volume = {358},
       number = {6370},
        pages = {1565-1570},
          doi = {10.1126/science.aap9580},
archivePrefix = {arXiv},
       eprint = {1710.05437},
 primaryClass = {astro-ph.HE},
       adsurl = {https://ui.adsabs.harvard.edu/abs/2017Sci...358.1565E}
}

@ARTICLE{2017PhRvL.119p1101A,
       author = {{Abbott}, B.~P. and {Abbott}, R. and {Abbott}, T.~D. and {Acernese}, F. and {Ackley}, K. and {Adams}, C. and {Adams}, T. and {Addesso}, P. and {Adhikari}, R.~X. and {Adya}, V.~B. and {Affeldt}, C. and {Afrough}, M. and {Agarwal}, B. and {Agathos}, M. and {Agatsuma}, K. and {Aggarwal}, N. and {Aguiar}, O.~D. and {Aiello}, L. and {Ain}, A. and {Ajith}, P. and {Allen}, B. and {Allen}, G. and {Allocca}, A. and {Altin}, P.~A. and {Amato}, A. and {Ananyeva}, A. and {Anderson}, S.~B. and {Anderson}, W.~G. and {Angelova}, S.~V. and {Antier}, S. and {Appert}, S. and {Arai}, K. and {Araya}, M.~C. and {Areeda}, J.~S. and {Arnaud}, N. and {Arun}, K.~G. and {Ascenzi}, S. and {Ashton}, G. and {Ast}, M. and {Aston}, S.~M. and {Astone}, P. and {Atallah}, D.~V. and {Aufmuth}, P. and {Aulbert}, C. and {AultONeal}, K. and {Austin}, C. and {Avila-Alvarez}, A. and {Babak}, S. and {Bacon}, P. and {Bader}, M.~K.~M. and {Bae}, S. and {Bailes}, M. and {Baker}, P.~T. and {Baldaccini}, F. and {Ballardin}, G. and {Ballmer}, S.~W. and {Banagiri}, S. and {Barayoga}, J.~C. and {Barclay}, S.~E. and {Barish}, B.~C. and {Barker}, D. and {Barkett}, K. and {Barone}, F. and {Barr}, B. and {Barsotti}, L. and {Barsuglia}, M. and {Barta}, D. and {Barthelmy}, S.~D. and {Bartlett}, J. and {Bartos}, I. and {Bassiri}, R. and {Basti}, A. and {Batch}, J.~C. and {Bawaj}, M. and {Bayley}, J.~C. and {Bazzan}, M. and {B{\'e}csy}, B. and {Beer}, C. and {Bejger}, M. and {Belahcene}, I. and {Bell}, A.~S. and {Berger}, B.~K. and {Bergmann}, G. and {Bernuzzi}, S. and {Bero}, J.~J. and {Berry}, C.~P.~L. and {Bersanetti}, D. and {Bertolini}, A. and {Betzwieser}, J. and {Bhagwat}, S. and {Bhandare}, R. and {Bilenko}, I.~A. and {Billingsley}, G. and {Billman}, C.~R. and {Birch}, J. and {Birney}, R. and {Birnholtz}, O. and {Biscans}, S. and {Biscoveanu}, S. and {Bisht}, A. and {Bitossi}, M. and {Biwer}, C. and {Bizouard}, M.~A. and {Blackburn}, J.~K. and {Blackman}, J. and {Blair}, C.~D. and {Blair}, D.~G. and {Blair}, R.~M. and {Bloemen}, S. and {Bock}, O. and {Bode}, N. and {Boer}, M. and {Bogaert}, G. and {Bohe}, A. and {Bondu}, F. and {Bonilla}, E. and {Bonnand}, R. and {Boom}, B.~A. and {Bork}, R. and {Boschi}, V. and {Bose}, S. and {Bossie}, K. and {Bouffanais}, Y. and {Bozzi}, A. and {Bradaschia}, C. and {Brady}, P.~R. and {Branchesi}, M. and {Brau}, J.~E. and {Briant}, T. and {Brillet}, A. and {Brinkmann}, M. and {Brisson}, V. and {Brockill}, P. and {Broida}, J.~E. and {Brooks}, A.~F. and {Brown}, D.~A. and {Brown}, D.~D. and {Brunett}, S. and {Buchanan}, C.~C. and {Buikema}, A. and {Bulik}, T. and {Bulten}, H.~J. and {Buonanno}, A. and {Buskulic}, D. and {Buy}, C. and {Byer}, R.~L. and {Cabero}, M. and {Cadonati}, L. and {Cagnoli}, G. and {Cahillane}, C. and {Calder{\'o}n Bustillo}, J. and {Callister}, T.~A. and {Calloni}, E. and {Camp}, J.~B. and {Canepa}, M. and {Canizares}, P. and {Cannon}, K.~C. and {Cao}, H. and {Cao}, J. and {Capano}, C.~D. and {Capocasa}, E. and {Carbognani}, F. and {Caride}, S. and {Carney}, M.~F. and {Carullo}, G. and {Casanueva Diaz}, J. and {Casentini}, C. and {Caudill}, S. and {Cavagli{\`a}}, M. and {Cavalier}, F. and {Cavalieri}, R. and {Cella}, G. and {Cepeda}, C.~B. and {Cerd{\'a}-Dur{\'a}n}, P. and {Cerretani}, G. and {Cesarini}, E. and {Chamberlin}, S.~J. and {Chan}, M. and {Chao}, S. and {Charlton}, P. and {Chase}, E. and {Chassande-Mottin}, E. and {Chatterjee}, D. and {Chatziioannou}, K. and {Cheeseboro}, B.~D. and {Chen}, H.~Y. and {Chen}, X. and {Chen}, Y. and {Cheng}, H. -P. and {Chia}, H. and {Chincarini}, A. and {Chiummo}, A. and {Chmiel}, T. and {Cho}, H.~S. and {Cho}, M. and {Chow}, J.~H. and {Christensen}, N. and {Chu}, Q. and {Chua}, A.~J.~K. and {Chua}, S.},
        title = "{GW170817: Observation of Gravitational Waves from a Binary Neutron Star Inspiral}",
      journal = {\prl},
         year = 2017,
        month = oct,
       volume = {119},
       number = {16},
          eid = {161101},
        pages = {161101},
          doi = {10.1103/PhysRevLett.119.161101},
archivePrefix = {arXiv},
       eprint = {1710.05832},
 primaryClass = {gr-qc},
       adsurl = {https://ui.adsabs.harvard.edu/abs/2017PhRvL.119p1101A}
}

@ARTICLE{2017ApJ...848L..12A,
       author = {{Abbott}, B.~P. and {Abbott}, R. and {Abbott}, T.~D. and {Acernese}, F. and {Ackley}, K. and {Adams}, C. and {Adams}, T. and {Addesso}, P. and {Adhikari}, R.~X. and {Adya}, V.~B. and {Affeldt}, C. and {Afrough}, M. and {Agarwal}, B. and {Agathos}, M. and {Agatsuma}, K. and {Aggarwal}, N. and {Aguiar}, O.~D. and {Aiello}, L. and {Ain}, A. and {Ajith}, P. and {Allen}, B. and {Allen}, G. and {Allocca}, A. and {Altin}, P.~A. and {Amato}, A. and {Ananyeva}, A. and {Anderson}, S.~B. and {Anderson}, W.~G. and {Angelova}, S.~V. and {Antier}, S. and {Appert}, S. and {Arai}, K. and {Araya}, M.~C. and {Areeda}, J.~S. and {Arnaud}, N. and {Arun}, K.~G. and {Ascenzi}, S. and {Ashton}, G. and {Ast}, M. and {Aston}, S.~M. and {Astone}, P. and {Atallah}, D.~V. and {Aufmuth}, P. and {Aulbert}, C. and {AultONeal}, K. and {Austin}, C. and {Avila-Alvarez}, A. and {Babak}, S. and {Bacon}, P. and {Bader}, M.~K.~M. and {Bae}, S. and {Baker}, P.~T. and {Baldaccini}, F. and {Ballardin}, G. and {Ballmer}, S.~W. and {Banagiri}, S. and {Barayoga}, J.~C. and {Barclay}, S.~E. and {Barish}, B.~C. and {Barker}, D. and {Barkett}, K. and {Barone}, F. and {Barr}, B. and {Barsotti}, L. and {Barsuglia}, M. and {Barta}, D. and {Barthelmy}, S.~D. and {Bartlett}, J. and {Bartos}, I. and {Bassiri}, R. and {Basti}, A. and {Batch}, J.~C. and {Bawaj}, M. and {Bayley}, J.~C. and {Bazzan}, M. and {B{\'e}csy}, B. and {Beer}, C. and {Bejger}, M. and {Belahcene}, I. and {Bell}, A.~S. and {Berger}, B.~K. and {Bergmann}, G. and {Bero}, J.~J. and {Berry}, C.~P.~L. and {Bersanetti}, D. and {Bertolini}, A. and {Betzwieser}, J. and {Bhagwat}, S. and {Bhandare}, R. and {Bilenko}, I.~A. and {Billingsley}, G. and {Billman}, C.~R. and {Birch}, J. and {Birney}, R. and {Birnholtz}, O. and {Biscans}, S. and {Biscoveanu}, S. and {Bisht}, A. and {Bitossi}, M. and {Biwer}, C. and {Bizouard}, M.~A. and {Blackburn}, J.~K. and {Blackman}, J. and {Blair}, C.~D. and {Blair}, D.~G. and {Blair}, R.~M. and {Bloemen}, S. and {Bock}, O. and {Bode}, N. and {Boer}, M. and {Bogaert}, G. and {Bohe}, A. and {Bondu}, F. and {Bonilla}, E. and {Bonnand}, R. and {Boom}, B.~A. and {Bork}, R. and {Boschi}, V. and {Bose}, S. and {Bossie}, K. and {Bouffanais}, Y. and {Bozzi}, A. and {Bradaschia}, C. and {Brady}, P.~R. and {Branchesi}, M. and {Brau}, J.~E. and {Briant}, T. and {Brillet}, A. and {Brinkmann}, M. and {Brisson}, V. and {Brockill}, P. and {Broida}, J.~E. and {Brooks}, A.~F. and {Brown}, D.~A. and {Brown}, D.~D. and {Brunett}, S. and {Buchanan}, C.~C. and {Buikema}, A. and {Bulik}, T. and {Bulten}, H.~J. and {Buonanno}, A. and {Buskulic}, D. and {Buy}, C. and {Byer}, R.~L. and {Cabero}, M. and {Cadonati}, L. and {Cagnoli}, G. and {Cahillane}, C. and {Calder{\'o}n Bustillo}, J. and {Callister}, T.~A. and {Calloni}, E. and {Camp}, J.~B. and {Canepa}, M. and {Canizares}, P. and {Cannon}, K.~C. and {Cao}, H. and {Cao}, J. and {Capano}, C.~D. and {Capocasa}, E. and {Carbognani}, F. and {Caride}, S. and {Carney}, M.~F. and {Casanueva Diaz}, J. and {Casentini}, C. and {Caudill}, S. and {Cavagli{\`a}}, M. and {Cavalier}, F. and {Cavalieri}, R. and {Cella}, G. and {Cepeda}, C.~B. and {Cerd{\'a}-Dur{\'a}n}, P. and {Cerretani}, G. and {Cesarini}, E. and {Chamberlin}, S.~J. and {Chan}, M. and {Chao}, S. and {Charlton}, P. and {Chase}, E. and {Chassande-Mottin}, E. and {Chatterjee}, D. and {Chatziioannou}, K. and {Cheeseboro}, B.~D. and {Chen}, H.~Y. and {Chen}, X. and {Chen}, Y. and {Cheng}, H. -P. and {Chia}, H. and {Chincarini}, A. and {Chiummo}, A. and {Chmiel}, T. and {Cho}, H.~S. and {Cho}, M. and {Chow}, J.~H. and {Christensen}, N. and {Chu}, Q. and {Chua}, A.~J.~K. and {Chua}, S. and {Chung}, A.~K.~W. and {Chung}, S. and {Ciani}, G.},
        title = "{Multi-messenger Observations of a Binary Neutron Star Merger}",
      journal = {\apjl},
         year = 2017,
        month = oct,
       volume = {848},
       number = {2},
          eid = {L12},
        pages = {L12},
          doi = {10.3847/2041-8213/aa91c9},
archivePrefix = {arXiv},
       eprint = {1710.05833},
 primaryClass = {astro-ph.HE},
       adsurl = {https://ui.adsabs.harvard.edu/abs/2017ApJ...848L..12A}
}

@ARTICLE{2017ApJ...848L..14G,
       author = {{Goldstein}, A. and {Veres}, P. and {Burns}, E. and {Briggs}, M.~S. and {Hamburg}, R. and {Kocevski}, D. and {Wilson-Hodge}, C.~A. and {Preece}, R.~D. and {Poolakkil}, S. and {Roberts}, O.~J. and {Hui}, C.~M. and {Connaughton}, V. and {Racusin}, J. and {von Kienlin}, A. and {Dal Canton}, T. and {Christensen}, N. and {Littenberg}, T. and {Siellez}, K. and {Blackburn}, L. and {Broida}, J. and {Bissaldi}, E. and {Cleveland}, W.~H. and {Gibby}, M.~H. and {Giles}, M.~M. and {Kippen}, R.~M. and {McBreen}, S. and {McEnery}, J. and {Meegan}, C.~A. and {Paciesas}, W.~S. and {Stanbro}, M.},
        title = "{An Ordinary Short Gamma-Ray Burst with Extraordinary Implications: Fermi-GBM Detection of GRB 170817A}",
      journal = {\apjl},
         year = 2017,
        month = oct,
       volume = {848},
       number = {2},
          eid = {L14},
        pages = {L14},
          doi = {10.3847/2041-8213/aa8f41},
archivePrefix = {arXiv},
       eprint = {1710.05446},
 primaryClass = {astro-ph.HE},
       adsurl = {https://ui.adsabs.harvard.edu/abs/2017ApJ...848L..14G}
}

@ARTICLE{2017ApJ...848L..15S,
       author = {{Savchenko}, V. and {Ferrigno}, C. and {Kuulkers}, E. and {Bazzano}, A. and {Bozzo}, E. and {Brandt}, S. and {Chenevez}, J. and {Courvoisier}, T.~J. -L. and {Diehl}, R. and {Domingo}, A. and {Hanlon}, L. and {Jourdain}, E. and {von Kienlin}, A. and {Laurent}, P. and {Lebrun}, F. and {Lutovinov}, A. and {Martin-Carrillo}, A. and {Mereghetti}, S. and {Natalucci}, L. and {Rodi}, J. and {Roques}, J. -P. and {Sunyaev}, R. and {Ubertini}, P.},
        title = "{INTEGRAL Detection of the First Prompt Gamma-Ray Signal Coincident with the Gravitational-wave Event GW170817}",
      journal = {\apjl},
         year = 2017,
        month = oct,
       volume = {848},
       number = {2},
          eid = {L15},
        pages = {L15},
          doi = {10.3847/2041-8213/aa8f94},
archivePrefix = {arXiv},
       eprint = {1710.05449},
 primaryClass = {astro-ph.HE},
       adsurl = {https://ui.adsabs.harvard.edu/abs/2017ApJ...848L..15S}
}

@ARTICLE{2017Sci...358.1556C,
       author = {{Coulter}, D.~A. and {Foley}, R.~J. and {Kilpatrick}, C.~D. and {Drout}, M.~R. and {Piro}, A.~L. and {Shappee}, B.~J. and {Siebert}, M.~R. and {Simon}, J.~D. and {Ulloa}, N. and {Kasen}, D. and {Madore}, B.~F. and {Murguia-Berthier}, A. and {Pan}, Y. -C. and {Prochaska}, J.~X. and {Ramirez-Ruiz}, E. and {Rest}, A. and {Rojas-Bravo}, C.},
        title = "{Swope Supernova Survey 2017a (SSS17a), the optical counterpart to a gravitational wave source}",
      journal = {Science},
         year = 2017,
        month = dec,
       volume = {358},
       number = {6370},
        pages = {1556-1558},
          doi = {10.1126/science.aap9811},
archivePrefix = {arXiv},
       eprint = {1710.05452},
 primaryClass = {astro-ph.HE},
       adsurl = {https://ui.adsabs.harvard.edu/abs/2017Sci...358.1556C}
}

@ARTICLE{2017ApJ...848L..17C,
       author = {{Cowperthwaite}, P.~S. and {Berger}, E. and {Villar}, V.~A. and {Metzger}, B.~D. and {Nicholl}, M. and {Chornock}, R. and {Blanchard}, P.~K. and {Fong}, W. and {Margutti}, R. and {Soares-Santos}, M. and {Alexander}, K.~D. and {Allam}, S. and {Annis}, J. and {Brout}, D. and {Brown}, D.~A. and {Butler}, R.~E. and {Chen}, H. -Y. and {Diehl}, H.~T. and {Doctor}, Z. and {Drout}, M.~R. and {Eftekhari}, T. and {Farr}, B. and {Finley}, D.~A. and {Foley}, R.~J. and {Frieman}, J.~A. and {Fryer}, C.~L. and {Garc{\'\i}a-Bellido}, J. and {Gill}, M.~S.~S. and {Guillochon}, J. and {Herner}, K. and {Holz}, D.~E. and {Kasen}, D. and {Kessler}, R. and {Marriner}, J. and {Matheson}, T. and {Neilsen}, Jr., E.~H. and {Quataert}, E. and {Palmese}, A. and {Rest}, A. and {Sako}, M. and {Scolnic}, D.~M. and {Smith}, N. and {Tucker}, D.~L. and {Williams}, P.~K.~G. and {Balbinot}, E. and {Carlin}, J.~L. and {Cook}, E.~R. and {Durret}, F. and {Li}, T.~S. and {Lopes}, P.~A.~A. and {Louren{\c{c}}o}, A.~C.~C. and {Marshall}, J.~L. and {Medina}, G.~E. and {Muir}, J. and {Mu{\~n}oz}, R.~R. and {Sauseda}, M. and {Schlegel}, D.~J. and {Secco}, L.~F. and {Vivas}, A.~K. and {Wester}, W. and {Zenteno}, A. and {Zhang}, Y. and {Abbott}, T.~M.~C. and {Banerji}, M. and {Bechtol}, K. and {Benoit-L{\'e}vy}, A. and {Bertin}, E. and {Buckley-Geer}, E. and {Burke}, D.~L. and {Capozzi}, D. and {Carnero Rosell}, A. and {Carrasco Kind}, M. and {Castander}, F.~J. and {Crocce}, M. and {Cunha}, C.~E. and {D'Andrea}, C.~B. and {da Costa}, L.~N. and {Davis}, C. and {DePoy}, D.~L. and {Desai}, S. and {Dietrich}, J.~P. and {Drlica-Wagner}, A. and {Eifler}, T.~F. and {Evrard}, A.~E. and {Fernandez}, E. and {Flaugher}, B. and {Fosalba}, P. and {Gaztanaga}, E. and {Gerdes}, D.~W. and {Giannantonio}, T. and {Goldstein}, D.~A. and {Gruen}, D. and {Gruendl}, R.~A. and {Gutierrez}, G. and {Honscheid}, K. and {Jain}, B. and {James}, D.~J. and {Jeltema}, T. and {Johnson}, M.~W.~G. and {Johnson}, M.~D. and {Kent}, S. and {Krause}, E. and {Kron}, R. and {Kuehn}, K. and {Nuropatkin}, N. and {Lahav}, O. and {Lima}, M. and {Lin}, H. and {Maia}, M.~A.~G. and {March}, M. and {Martini}, P. and {McMahon}, R.~G. and {Menanteau}, F. and {Miller}, C.~J. and {Miquel}, R. and {Mohr}, J.~J. and {Neilsen}, E. and {Nichol}, R.~C. and {Ogando}, R.~L.~C. and {Plazas}, A.~A. and {Roe}, N. and {Romer}, A.~K. and {Roodman}, A. and {Rykoff}, E.~S. and {Sanchez}, E. and {Scarpine}, V. and {Schindler}, R. and {Schubnell}, M. and {Sevilla-Noarbe}, I. and {Smith}, M. and {Smith}, R.~C. and {Sobreira}, F. and {Suchyta}, E. and {Swanson}, M.~E.~C. and {Tarle}, G. and {Thomas}, D. and {Thomas}, R.~C. and {Troxel}, M.~A. and {Vikram}, V. and {Walker}, A.~R. and {Wechsler}, R.~H. and {Weller}, J. and {Yanny}, B. and {Zuntz}, J.},
        title = "{The Electromagnetic Counterpart of the Binary Neutron Star Merger LIGO/Virgo GW170817. II. UV, Optical, and Near-infrared Light Curves and Comparison to Kilonova Models}",
      journal = {\apjl},
         year = 2017,
        month = oct,
       volume = {848},
       number = {2},
          eid = {L17},
        pages = {L17},
          doi = {10.3847/2041-8213/aa8fc7},
archivePrefix = {arXiv},
       eprint = {1710.05840},
 primaryClass = {astro-ph.HE},
       adsurl = {https://ui.adsabs.harvard.edu/abs/2017ApJ...848L..17C}
}

@ARTICLE{2017Natur.551...80K,
       author = {{Kasen}, Daniel and {Metzger}, Brian and {Barnes}, Jennifer and {Quataert}, Eliot and {Ramirez-Ruiz}, Enrico},
        title = "{Origin of the heavy elements in binary neutron-star mergers from a gravitational-wave event}",
      journal = {\nat},
         year = 2017,
        month = nov,
       volume = {551},
       number = {7678},
        pages = {80-84},
          doi = {10.1038/nature24453},
archivePrefix = {arXiv},
       eprint = {1710.05463},
 primaryClass = {astro-ph.HE},
       adsurl = {https://ui.adsabs.harvard.edu/abs/2017Natur.551...80K}
}

@ARTICLE{2018ApJ...856L..18M,
       author = {{Margutti}, R. and {Alexander}, K.~D. and {Xie}, X. and {Sironi}, L. and {Metzger}, B.~D. and {Kathirgamaraju}, A. and {Fong}, W. and {Blanchard}, P.~K. and {Berger}, E. and {MacFadyen}, A. and {Giannios}, D. and {Guidorzi}, C. and {Hajela}, A. and {Chornock}, R. and {Cowperthwaite}, P.~S. and {Eftekhari}, T. and {Nicholl}, M. and {Villar}, V.~A. and {Williams}, P.~K.~G. and {Zrake}, J.},
        title = "{The Binary Neutron Star Event LIGO/Virgo GW170817 160 Days after Merger: Synchrotron Emission across the Electromagnetic Spectrum}",
      journal = {\apjl},
         year = 2018,
        month = mar,
       volume = {856},
       number = {1},
          eid = {L18},
        pages = {L18},
          doi = {10.3847/2041-8213/aab2ad},
archivePrefix = {arXiv},
       eprint = {1801.03531},
 primaryClass = {astro-ph.HE},
       adsurl = {https://ui.adsabs.harvard.edu/abs/2018ApJ...856L..18M}
}

@ARTICLE{2018ApJ...863L..18A,
       author = {{Alexander}, K.~D. and {Margutti}, R. and {Blanchard}, P.~K. and {Fong}, W. and {Berger}, E. and {Hajela}, A. and {Eftekhari}, T. and {Chornock}, R. and {Cowperthwaite}, P.~S. and {Giannios}, D. and {Guidorzi}, C. and {Kathirgamaraju}, A. and {MacFadyen}, A. and {Metzger}, B.~D. and {Nicholl}, M. and {Sironi}, L. and {Villar}, V.~A. and {Williams}, P.~K.~G. and {Xie}, X. and {Zrake}, J.},
        title = "{A Decline in the X-Ray through Radio Emission from GW170817 Continues to Support an Off-axis Structured Jet}",
      journal = {\apjl},
         year = 2018,
        month = aug,
       volume = {863},
       number = {2},
          eid = {L18},
        pages = {L18},
          doi = {10.3847/2041-8213/aad637},
archivePrefix = {arXiv},
       eprint = {1805.02870},
 primaryClass = {astro-ph.HE},
       adsurl = {https://ui.adsabs.harvard.edu/abs/2018ApJ...863L..18A}
}

@ARTICLE{2020ApJ...890L..24Z,
       author = {{Zhang}, Bing},
        title = "{Fast Radio Bursts from Interacting Binary Neutron Star Systems}",
      journal = {\apjl},
         year = 2020,
        month = feb,
       volume = {890},
       number = {2},
          eid = {L24},
        pages = {L24},
          doi = {10.3847/2041-8213/ab7244},
archivePrefix = {arXiv},
       eprint = {2002.00335},
 primaryClass = {astro-ph.HE},
       adsurl = {https://ui.adsabs.harvard.edu/abs/2020ApJ...890L..24Z}
}

@ARTICLE{2020ApJ...893L...6M,
       author = {{Most}, Elias R. and {Philippov}, Alexander A.},
        title = "{Electromagnetic Precursors to Gravitational-wave Events: Numerical Simulations of Flaring in Pre-merger Binary Neutron Star Magnetospheres}",
      journal = {\apjl},
         year = 2020,
        month = apr,
       volume = {893},
       number = {1},
          eid = {L6},
        pages = {L6},
          doi = {10.3847/2041-8213/ab8196},
archivePrefix = {arXiv},
       eprint = {2001.06037},
 primaryClass = {astro-ph.HE},
       adsurl = {https://ui.adsabs.harvard.edu/abs/2020ApJ...893L...6M}
}
\bibliographystyle{aasjournal}



\end{document}